\documentclass[conference]{IEEEtran}
\usepackage{longtable, graphicx}
\usepackage{placeins} % This package provides the %\FloatBarrier command
\usepackage{longtable}
\usepackage{parskip, indentfirst}
\usepackage{etoc} % for local table of contents
\usepackage{amsmath,amssymb,amsfonts,amsthm}
\usepackage[linesnumbered,ruled,vlined]{algorithm2e}

\usepackage{mathtools, braket, dsfont, graphicx, lipsum, enumitem}
\usepackage{textcomp, float, etoc, bibentry, multirow, tabularx}
\usepackage{nicefrac}       % compact symbols for 1/2, etc.
\usepackage{microtype}      % microtypography
\usepackage{nomencl}                    % produces a nomenclature
\usepackage{float}                      % figure floats
\usepackage{fancyhdr}                   % fancy headers and footers
\usepackage{url}                        % nicely format url breaks
\usepackage{relsize}                    % font sizing hierarchy
\usepackage{booktabs}                   % professional looking tables
\usepackage{mathrsfs}                   % additional math scripts
\usepackage{pdflscape}			% to produce landscape pages if necessary
\usepackage{tikz}
\usetikzlibrary{quantikz}
\usepackage[numbers]{natbib}
\usepackage[inactive]{srcltx}		 	% necessary to use forward and inverse searching in DVI
\usepackage[config, labelfont={bf}]{caption,subfig} % nice sub figures
\usepackage[normalem]{ulem}
\usepackage[english]{babel}
\usepackage[utf8]{inputenc} % allow utf-8 input
\usepackage[T1]{fontenc}    % use 8-bit T1 fonts
\usepackage{hyperref}
\hypersetup{
  colorlinks=true, 
  allcolors=blue,
  pdfauthor={...},
  pdftitle={...},
  pdfsubject={...},
}
\usepackage{xstring}
\usepackage{xspace}
\newcolumntype{P}[1]{>{\raggedright\arraybackslash}p{#1}}
\newcommand{\added}[1]{{\color{black}#1}}

\newcommand{\xrm}{\textrm{x}}

\newcommand{\low}[1]{{{}_{\text{#1}}}}
\newcommand{\BV}{Bernstein-Vazirani\xspace}
\newcommand{\pec}{probabilistic error cancellation\xspace}
\newcommand{\nsn}{non-stationary noise\xspace}
\newcommand{\ns}{non-stationary\xspace}
\newcommand{\SPAM}{state preparation and measurement\xspace}

\newcommand{\kolkata}{ibm\_kolkata\xspace}
\newcommand{\Tr}{\textrm{Tr}}
\newcommand{\sop}{\textrm{super-operator}}
\newcommand{\superc}{\textrm{superconducting}}

\def\BibTeX{{\rm B\kern-.05em{\superc~i\kern-.025em b}\kern-.08em
    T\kern-.1667em\lower.7ex\hbox{E}\kern-.125emX}}

\graphicspath{
{figures/}
}

\title{Improving probabilistic error cancellation\\ in the presence of non-stationary noise
}
\author{
\IEEEauthorblockN{Samudra~Dasgupta$^{1,2^*}$ and Travis S.~Humble$^{1,2^\dagger}$}
\IEEEauthorblockA{\textit{$^1$Quantum Science Center, Oak Ridge National Laboratory, Oak Ridge, Tennessee, USA}\\
\textit{$^{2}$Bredesen Center, University of Tennessee, Knoxville, USA}
}
\thanks{
This manuscript has been authored by UT-Battelle, LLC under Contract No. DE-AC05-00OR22725 with the U.S. Department of Energy. The United States Government retains and the publisher, by accepting the article for publication, acknowledges that the United States Government retains a non-exclusive, paid-up, irrevocable, worldwide license to publish or reproduce the published form of this manuscript, or allow others to do so, for United States Government purposes. The Department of Energy will provide public access to these results of federally sponsored research in accordance with the DOE Public Access Plan (https://www.energy.gov/doe-public-access-plan).
}
}

\IEEEoverridecommandlockouts
\begin{document}
\maketitle

\begin{abstract}
We investigate the stability of probabilistic error cancellation (PEC) outcomes in the presence of non-stationary noise, which is an obstacle to achieving accurate observable estimates. Leveraging Bayesian methods, we design a strategy to enhance PEC stability and accuracy. Our experiments using a 5-qubit implementation of the Bernstein-Vazirani algorithm and conducted  on the \kolkata device reveal a 42\% improvement in accuracy and a 60\% enhancement in stability compared to non-adaptive PEC. These results underscore the importance of adaptive estimation processes to effectively address non-stationary noise, vital for advancing PEC utility.
\end{abstract}


%\tableofcontents

\section{Introduction}
Error-resilient quantum computing holds great promise, offering significant advancements over conventional computing. Once realized, it is expected to super-polynomially reduce execution time, energy consumption, and memory storage needs compared to conventional state-of-the-art computers~\cite{humble2019quantum}. The potential impact of error-resilient quantum computing includes revolutionizing scientific applications such as simulating many-body quantum systems~\cite{browaeys2020many}, solving large-scale optimization problems~\cite{montanaro2016quantum}, efficiently sampling high-dimensional probability distributions~\cite{shang2015monte}, factorizing large integers, and enhancing the security of communication networks~\cite{espitia2021role}. Consequently, this technology is expected to be disruptive to sectors such as telecommunications, cyber-security, pharmaceuticals, logistics, supply chain management, artificial intelligence, and materials science~\cite{how2023business}. 

The practical realization of quantum computing has witnessed rapid advancements~\cite{roadmap}, with quantum devices now operating as systems with hundreds of interacting qubits. However, these real quantum devices~\cite{roth2021introduction} are noisy~\cite{preskill2019quantum}, and practical efforts to realize a quantum computer introduce various noise processes like decay, de-coherence~\cite{kandala2019error}, environmental coupling, intra-register cross-talk~\cite{parrado2021crosstalk, fang2022crosstalk}, and leakage from computational space~\cite{vittal2023eraser}. Physical operations like quantum gates and measurements rely on electromagnetic fields susceptible to pulse distortion, attenuation, jitter, and drift, which further increase noise~\cite{kliesch2021theory,PRXQuantum.3.020344}. Imperfections in thermodynamic controls (such as cryogenic cooling, magnetic shielding, vibration suppression, and imperfect vacuum chambers) can disturb the operating conditions of the quantum computer~\cite{blume2020modeling, blume2010optimal}. 

These lead to computational errors that make it essential to address noise and implement error mitigation strategies~\cite{bharti2022noisy} to improve the accuracy of quantum outcome~\cite{ferracin2021experimental}.

%%%%%%%%%%%%%%%%%%%%%%%%%%%%%%%%%%%%%%%%%%%%%%%%%%%%%%%%%%%%%%%%%%%%%%%%
\begin{figure}[htbp]
\centering
\includegraphics[width=\linewidth]{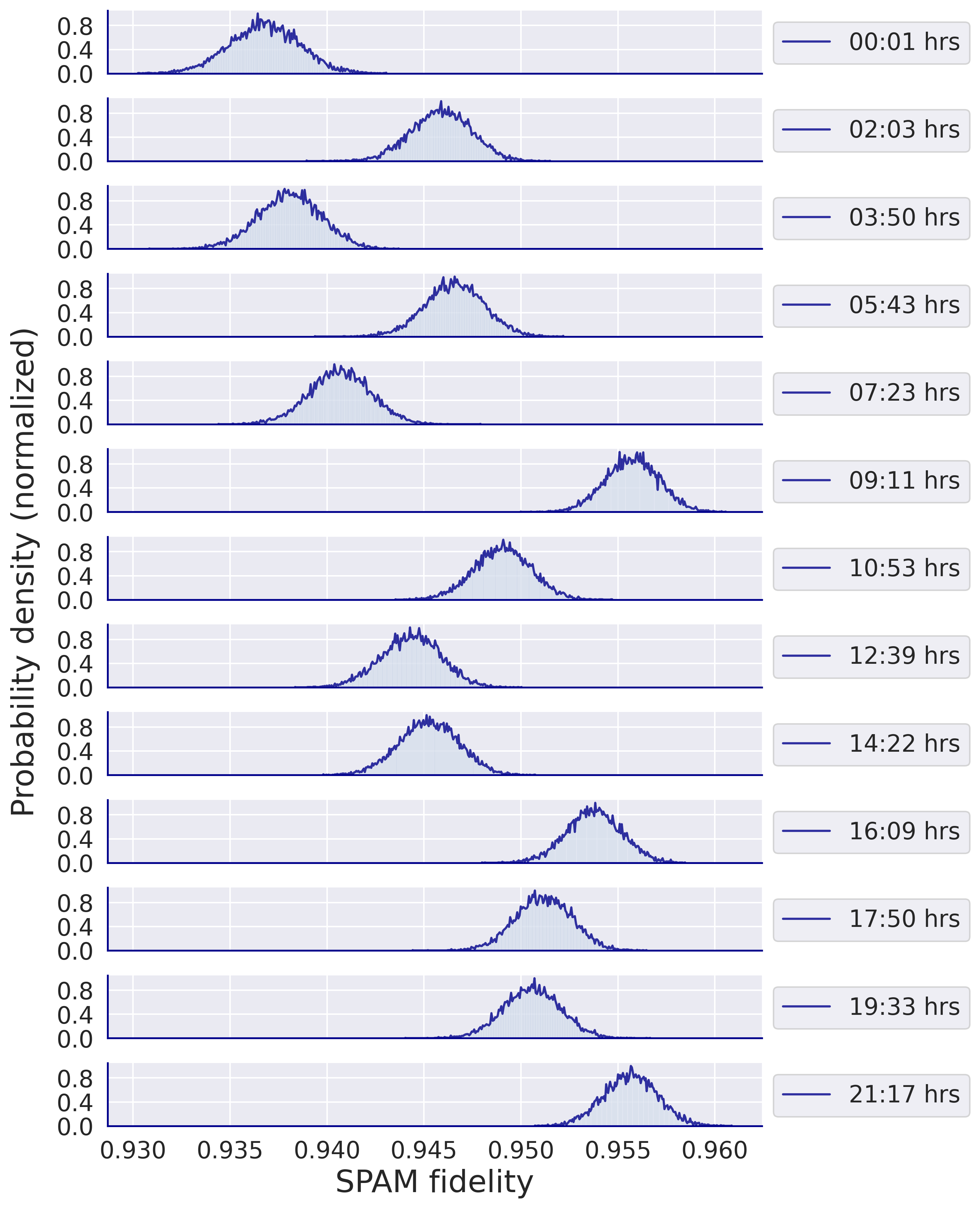}
\caption{Non-stationary distribution functions of the \SPAM (SPAM) fidelity for qubit 2 on \kolkata~\superc~device collected on Jan 15, 2024.}
\label{fig:tv_noise_f2}
\end{figure}
%%%%%%%%%%%%%%%%%%%%%%%%%%%%%%%%%%%%%%%%%%%%%%%%%%%%%%%%%%%%%%%%%%%%%%%%

%\subsubsection{Why does noise vary with time?}
Contemporary quantum computers are not only noisy but they also exhibit non-stationarity. Non-stationary noise processes~\cite{thorbeck2023two, mcewen2022resolving, etxezarreta2021time, muller2015interacting, klimov2018fluctuations} in \superc~qubits~\cite{martinis2015qubit} are well-studied. For example, \SPAM (SPAM) fidelities have been observed to fluctuate significantly, showing more than 25\% deviation from their long-term average~\cite{dasgupta2023reliability}. 
Similarly, the fidelity of CNOT gates have been noted to change by over 40\% within similar time frames~\cite{dasgupta2021stability}. 
Moreover, the qubit relaxation time ($T_1$) has been observed to experience fluctuations of up to 400\% in just 30 minutes~\cite{dasgupta2020characterizing}. 
Likewise, de-phasing time ($T_2$) has been recorded to vary by over 50\% within an hour~\cite{carroll2021dynamics, mcrae2021reproducible}. 

\added{The non-stationarity observed in contemporary \superc~quantum computers stems from two primary sources linked to material defects: impurities within the material and ionization induced by cosmic rays.} 
It is theorized that fluctuating two-level systems, possibly stemming from certain oxides on the superconductor's surface, contribute to non-stationarity~\cite{muller2015interacting, klimov2018fluctuations}.
Additionally, cosmic rays~\cite{xu2022distributed, mcewen2022resolving} contribute by ionizing the substrate upon impact, leading to the emission of high-energy phonons, which in turn triggers a burst of quasi-particles. These quasi-particles disrupt qubit coherence across the device. It has been shown that quantum computers can experience catastrophic errors in multi-qubit registers approximately every 10 seconds due to cosmic rays originating from outer space~\cite{mcewen2022resolving}. Studies that address non-stationary noise in \superc~quantum computers include 
investigations on output reproducibility~\cite{proctor2020detecting},  
noise modeling~\cite{etxezarreta2021time}, 
tracking the \ns~profile of quantum noise~\cite{danageozian2022noisy}, and 
quantum error mitigation using continuous control~\cite{majumder2020real}.

Quantum error mitigation is a set of techniques that employs statistical tools from estimation theory to reduce the impact of noise in quantum computations without directly correcting the quantum state~\cite{bharti2022noisy}. Such techniques can become vulnerable to errors stemming from over or under estimation of noise due to the presence of non-stationarity~\cite{kim2024error, henao2023adaptive}. 

A quantum noise channel~\cite{nielsen2002quantum} can be described as a stochastic process, allowing for the continuous update of its estimated characteristics in response to varying noise conditions. Such an approach treats the channel as a time-varying random variable. Consider Fig.~\ref{fig:tv_noise_f2} which shows the \SPAM (SPAM) fidelity for the second register element on \kolkata~device on Jan 15, 2024. The probability density is clearly changing with time, even though the variance stays consistent. In light of such non-stationary data, a single-qubit SPAM noise channel can be described by the model: $\mathcal{E}(\rho) = f(t) \rho + [1-f(t)] \mathds{X} \rho \mathds{X}$, where $\rho$ represents the single-qubit density matrix, $\mathds{X} = 
\begin{pmatrix}
0 & 1 \\
1 & 0
\end{pmatrix}$ denotes the Pauli-X matrix, and $f(t)$ denotes the SPAM fidelity drawn from a time-dependent distribution. From Fig.~\ref{fig:tv_noise_f2}, non-stationarity of noise in this \superc~device is apparent within 24 hours. The noise parameters estimated during device re-calibration quickly becomes outdated, compromising the accuracy of noise channel information essential for mitigation. Thus re-calibration alone is insufficient and there is a need for adaptive error mitigation techniques that can function in between calibration intervals in the face of changing noise conditions.

This study focuses on probabilistic error cancellation (PEC)~\cite{temme2017error} in the presence of non-stationary noise. PEC is a quantum error mitigation approach that aims to construct unbiased estimates of the means of quantum observables from noisy observations. 
Effective implementation requires an accurate noise characterization~\cite{torlai2020quantum}. 
For example, 
learning correlated noise channels in large quantum circuits on a \superc~quantum processor has proven to be difficult~\cite{van2023probabilistic}. 
Yet, leveraging sparse noise models, PEC has been successful in estimating the mean of observables in circuits comprising 2,880 CNOT gates, executed on a 127-qubit noisy \superc~processor - a task that conventional brute-force computing could not match~\cite{kim2023evidence}.

We demonstrate that adaptive probabilistic error cancellation, which views quantum noise channels as evolving random variables, outperforms its non-adaptive counterpart in devices subject to non-stationary noise. To achieve this aim, we will make use of a 5-qubit implementation of the \BV algorithm~\cite{bernstein1993quantum}. 

The manuscript is organized as follows. 
In Sec.~\ref{sec:background}, we provide background for \pec (PEC). 
Sec.~\ref{sec:adaptPEC} develops a Bayesian~\cite{lukens2020practical, zheng2020bayesian} approach for adapting the method of \pec to \nsn. It also sets up a performance evaluation framework for the accuracy and stability of PEC results. 
Sec.~\ref{sec:PEC_under_Pauli} presents a numerical validation of adaptive probabilistic error cancellation using a 5-qubit implementation of the \BV algorithm, where we treat the noise parameters (qubit-specific \SPAM (SPAM) fidelities and depolarizing parameters characterizing noise in CNOT gate) as \ns~random variables.
In Sec.~\ref{sec:real_data}, we present the results of experiments conducted on a real, noisy quantum device to test the adaptive PEC algorithm. Concluding remarks are provided in Sec.~\ref{sec:conclusion}.
\section{Background}\label{sec:background}
%\subsection{PEC}\label{sec:pec-in-brief}
In this section, we provide background for the quantum error mitigation method called probabilistic error cancellation (PEC)~\cite{temme2017error, van2023probabilistic, kim2023evidence} \added{which aims to mitigate errors by approximating the noiseless mean of the observable as a weighted sum of noisy observables. }
% It is motivated from linear algebra, where any vector can be represented as a linear combination of basis vectors. In the PEC context, the vectors are the  super-operators, and the method involves expressing a desired super-operator as a combination of basis super-operators in the super-operator space. Specifically, the PEC method involves designing a set of modified quantum circuits, where each modification involves integrating  noisy, basis operations available into the original circuit. By executing these modified circuits and then combining their results using predetermined weights, the approach aims to approximate the outcome that would have been achieved with a noiseless computer.} 
We use calligraphic symbols to denote {\sop}s acting on density matrices ($\rho$): 
\begin{equation}
\mathcal{G} \rho = G\rho G^\dagger.
\end{equation}
where $\mathcal{G}$ is the {\sop} and $G$ is a unitary quantum operator. 
For example, if $G$ denotes the CNOT operator, then $\mathcal{G}$ is the {\sop} for the CNOT operation.

Typically we do not have access to a noiseless implementation of $\mathcal{G}$. Let $\tilde{\mathcal{G}}$ denote the \sop~corresponding to the available, noisy implementation of $\mathcal{G}$. We have access to other synthesized implementations of $\tilde{\mathcal{G}}$ by subjecting $\tilde{\mathcal{G}}$ to basis operations available on a noisy device. 
\added{Originally, the noisy basis set was specified as the set of native gates that a quantum computer could implement~\cite{temme2017error}. However, with advancements in \superc~quantum computers, the noise associated with single-qubit Pauli operators has become negligible~\cite{van2023probabilistic, bravyi2021mitigating}. Consequently, Pauli operators, which might be a composition of multiple native gates, can be employed as the basis set~\cite{mitiq-qpr}.} 
We denote by $\{\tilde{\mathcal{G}_k}\}$ the set of all noisy {\sop}s, with $k=0$ denoting the noisy implementation of $\mathcal{G}$ without additional operations from the basis set.

As an example, consider a 2-qubit quantum gate $G$. The set of noisy {\sop}s composed under a Pauli channel assumption are given by: $\{ \mathcal{P} \circ \tilde{\mathcal{G}} \}$ where 
$\mathcal{P} \circ \tilde{\mathcal{G}} (\rho) = \mathcal{P} ( \tilde{\mathcal{G}} (\rho) )$ and 
$\mathcal{P}(\cdot) \equiv [\mathds{P}_0 \otimes \mathds{P}_1](\cdot)[\mathds{P}_0 \otimes \mathds{P}_1]$. 
Here $0$ and $1$ refer to qubit $0$ and qubit $1$ respectively and $\mathds{P}_0, \mathds{P}_1$ are picked from the set of Pauli operators $\{ \mathds{I}, \mathds{X},\mathds{Y},\mathds{Z}\}$. 
%When $P_0 = P_1 = \mathds{I}$, we retrieve back $\tilde{\mathcal{G}}$. 
Thus, by varying $P_0$ and $P_1$, we obtain the set $\{ \mathcal{P} \circ \tilde{\mathcal{G}} \}$ which forms a basis $\{\tilde{\mathcal{G}_k}\}_{k=0}^{15}$ that spans the \sop~space. In this 2-qubit example, the map for $k$ is derived from the Cartesian product of $\{ \mathds{I}, \mathds{X},\mathds{Y},\mathds{Z}\} \times \{ \mathds{I}, \mathds{X},\mathds{Y},\mathds{Z}\}$, with the sequence of this ordered set determining the value of $k$.

In general, the {\sop} $\mathcal{G}$ can be expressed as a linear combination of the basis {\sop}s:
\begin{equation}
\mathcal{G} = \sum\limits_{k=0}^{N_p-1} \eta_\low{k} \tilde{\mathcal{G}}_k,
\label{eq:pec}
\end{equation}
where $N_p$ is the dimension of the \sop~space. The PEC coefficients ${ \eta_\low{k} }$ in the linear combination are determined either analytically, under a noise model assumption for single and two-qubit gates, which can be extended to larger circuits, or numerically, by minimizing the one-norm between high-dimensional matrices~\cite{mitiq-qpr, temme2017error}. We employ the analytical approach.

%For a Pauli noise model, the number of terms, denoted as $N_p$, equals $4^n$, where $n$ is the number of qubits. 
%The factor $4$ arises from the dimensionality of the set of single-qubit Pauli matrices $\{\mathds{I}, \mathds{X}, \mathds{Y}, \mathds{Z}\}$.

%By definition, the noisy {\sop}s are perfectly implementable. 

The circuits corresponding to the {\sop}s $\tilde{\mathcal{G}}_k$ in Eqn.~\ref{eq:pec} are constructed by subjecting each of the gates in the quantum circuit for $\tilde{\mathcal{G}}$ to operations from the noisy basis set. 
% (e.g. $\mathds{I}, \mathds{X}, \mathds{Y}, \mathds{Z}$, CNOT). 
If we execute these noisy circuits and collect the mean of the observable, then, from Eqn.~\ref{eq:pec}, we recover the ideal noiseless mean of an observable $\braket{\mathcal{O}}$ as:
\begin{equation}
\braket{\mathcal{O}} = \Tr\left[ \mathcal{O} \mathcal{G} \rho \right] = \Tr\left[ \mathcal{O} \sum\limits_k \eta_\low{k} \tilde{\mathcal{G}}_k \rho \right]
\label{eq:pec_basic}
\end{equation}
where $\rho$ is the input density matrix to the circuit. Thus, an observable $\braket{\mathcal{O}}$ may be estimated as a weighted sum of the mean of the observables from the noisy circuits. 

Eqn.~\ref{eq:pec_basic} can be re-written as:
\begin{equation}
\braket{\mathcal{O}} = \gamma \sum\limits_k \text{sgn}(\eta_\low{k}) \mathds{Q}_\low{k}  \Tr \left[
\mathcal{O} \tilde{\mathcal{G}_\low{k}}\rho
\right]
\label{eq:pec_estimator}
\end{equation}
with $\gamma = \sum |\eta_\low{k}|$ and $\mathds{Q}_\low{k} = |\eta_\low{k}|/\gamma$. 
The sign function, \( \text{sgn}(\cdot) \), returns +1 for positive inputs, -1 for negative inputs, and 0 for an input of 0.

Note that the set $\{\mathds{Q}_k\}$ forms a valid probability distribution because all its elements are positive and sum to 1. 
However, $\{\mathds{Q}_k\}$ is termed a quasi-probability distribution (QPD)~\cite{temme2017error, mitiq-qpr}. To see this, consider a random integer $K \in \{0, \cdots N_p-1\}$ which follows the probability distribution function denoted by $\mathds{Q}_k$. This random integer $K$ can be mapped one-to-one to the random variable 
$\text{sgn}(\eta_\low{K}) \Tr \left[  \mathcal{O}  \tilde{\mathcal{G}_\low{K}}\rho \right]$, 
which inherits the probability distribution function $\mathds{Q}_k$. By repeatedly sampling the random variable $K$, we realize a set of random integers represented as $\{\mathfrak{m}\}$. 
For each specific $\mathfrak{m}$ that is realized, \added{we execute the corresponding noisy quantum circuit $\tilde{\mathcal{G}_\mathfrak{m}}$ multiple times} and obtain a set of noisy means of observables denoted by the set $\left\{ \Tr \left[ \mathcal{O} \tilde{\mathcal{G}_\mathfrak{m}}\rho \right] \right\}$. 

When computing the average over the set of noisy means of observables $\left\{ \Tr \left[ \mathcal{O} \tilde{\mathcal{G}_\mathfrak{m}}\rho \right] \right\}$, we need to 
adjust the sign of each element of the set by the sign of $\eta_\low{m}$ and scale it by $\gamma$.
This average then converges to the mean of the observable from a noiseless gate $\mathcal{G}$, in the asymptotic limit of a large number of repeated samplings of the random variable $K$. The need for adjustment by the sign function leads us to denote $\mathds{Q}_k$ as quasi-probabilities.% for the observables obtained from $\tilde{\mathcal{G}_\mathfrak{m}}$.

\added{Achieving an accuracy of \(O(\epsilon)\) using the empirical mean of the random variable 
\(\gamma \text{sgn}(\eta_\low{K}) \Tr \left[ \mathcal{O} \tilde{\mathcal{G}_\low{K}}\rho \right]\) 
(which is an unbiased estimator of \(\braket{\mathcal{O}}\))
requires \(O(\gamma/ \epsilon)^2\) PEC circuit samples and the result has variance of order \(O(\gamma^2)\) \cite{temme2017error}.}
\section{Adaptive PEC}\label{sec:adaptPEC}
%\subsection{Adaptive PEC}
Estimating channel noise parameters is crucial for determining the PEC coefficients $\{\eta_\low{k}\}$. 
However, the non-stationary nature of noise, along with drift and latency in characterization, complicates this task. This in turn makes it difficult to accurately assess the PEC coefficients. In this section, we demonstrate how adaptive parameter estimation can be applied to PEC.

The parameters characterizing the noise during idle time (such as qubit decoherence time~\cite{burnett2019decoherence}) and quantum operations (such as CNOT fidelity~\cite{dasgupta2023reliability}) exhibit random non-stationary behavior in some hardware. Estimating non-stationary stochastic processes is challenging and their predictive value is limited because the patterns identified from historical data may not reliably indicate future behavior, making it difficult to discern underlying trends. 
A model that is effective at one time point can become inaccurate at another. 

However, we can utilize intermittent incremental measurements from quantum circuits to devise a Bayesian~\cite{lukens2020practical, zheng2020bayesian, gordon1993novel, kotecha2003gaussian} update for the current state of the device noise:
\begin{equation}
%\Pr(\textbf{x} | \text{data} ) \propto \frac{\Pr(\text{data}|\textbf{x}) \Pr(\textbf{x})}{ \Pr(\text{data})}
\Pr(\textbf{x} | \text{data} ) \propto \Pr(\text{data}|\textbf{x}) \Pr(\textbf{x})
\label{eq:basic_bayes}
\end{equation}
where 
data refers to measurements obtained from circuit execution, 
$\textbf{x}$ denotes the multi-dimensional vector of parameters characterizing the device noise (such as connection-specific CNOT fidelity and qubit-specific SPAM fidelities), 
$\Pr(\textbf{x} | \text{data} )$ is the posterior noise distribution, 
$\Pr(\text{data}|\textbf{x} )$ is the likelihood, 
and $\Pr(\textbf{x})$ is the prior noise distribution. 
%and 
%$\Pr(\text{data})$ is a normalizing constant that does not depend on the parameters of the probability distribution function.
We use this to update $\{\eta_\low{k}\}$ via updates to model-specific parameters. 

Obtaining the posterior distribution for $\xrm$ is a two-step process: first, we estimate the posterior for uncorrelated parameters, and second, for correlated parameters in the underlying noise model.\\
\subsection{Uncorrelated parameters}
Uncorrelated parameters refer to those model parameters for which there exist datasets (such as the partial trace of the observable on a single qubit) where the observed data is modeled by error due to only one noise parameter. 
Here, the analysis is simpler as univariate priors can be used. 
%For example, symmetric readout noise is a one-parameter model.
For example, if we disregard noise from single-qubit rotations, then the measurements obtained from any qubit that was not subjected to entangling operations, can be described by a one-parameter model under symmetric SPAM noise model. 

Let $\mathcal{D}_q = \{b_q(0), \cdots, b_q(L-1)\} $ represent a dataset of $L$ samples obtained for qubit $q$ after measurement in the computational basis. 
Each $b_q(l)$ is a single-bit measured after the $l$-th execution of $\tilde{\mathcal{G}}_0$. We can adopt a beta distribution as the prior for the SPAM fidelity $f_q$ because the beta distribution is well-suited for values restricted to the [0,1] interval and can effectively accommodate the experimentally observed unimodal and skewed density as seen in Fig.~\ref{fig:tv_noise_f2}. The beta distribution's flexibility allows for an accurate fit to these characteristics. The Beta distribution with parameters $\alpha_q, \beta_q$ is given by $\text{Beta}(f_q; \alpha_q, \beta_q) = f_q^{\alpha_q-1}(1-f_q)^{\beta_q-1} / B(\alpha_q,\beta_q)$ where 
the normalizing denominator $B(\alpha_q,\beta_q)$ is the Beta function defined as $B(m,n) = \Gamma(m)\Gamma(n)/\Gamma(m+n)$, and $\Gamma(\cdot)$ is the gamma function $\Gamma(y) = \int\limits_0^\infty t^{y-1} e^{-t}dt$, defined for any positive y.

The prior for the mean $\mu_q^\text{prior}$ and variance $v_q^\text{prior}$ of the SPAM fidelity $f_q$ can be derived from historical characterization data (e.g. using data post calibration). If such data is not accessible, we can obtain the starting point from a small perturbation to the ideal value. 
%When presented with new information, the Bayesian inference method adjusts the priors.
The prior parameters are then obtained as: $\alpha_q^\text{prior} = \mu_q^\text{prior} \left[ \mu_q^\text{prior}(1-\mu_q^\text{prior}) / v_q^\text{prior} - 1\right]$ and $\beta_q^\text{prior} = (1-\mu_q^\text{prior}) \left[ \mu_q^\text{prior}(1-\mu_q^\text{prior}) / v_q^\text{prior} -1\right]$.

The likelihood is obtained as:
\begin{equation}
%\begin{split}
\mathcal{L}(f_q)
%=& \text{Pr}\left[ \mathcal{D}_q | f_q \right] \\
%=& \prod\limits_{l=0}^{L-1} \text{Pr} \left[ b_q(l) | f_q \right] \\
%=& \prod\limits_{l=0}^{L-1} f_q^{\delta_0[b_q(l)]} (1-f_q)^{1-\delta_0[b_q(l)]} \\
%=& f_q^{\sum\limits_l \delta_0[b_q(l)]} (1-f_q)^{L-\sum\limits_l \delta_0[b_q(l)]} \\
= f_q^{C_0[\mathcal{D}_q]} (1-f_q)^{L-C_0[\mathcal{D}_q]} / B^L(\alpha_q, \beta_q)
%\end{split}
\end{equation}
where $C_0[\mathcal{D}_q]$ counts the number of 0's in the dataset $\mathcal{D}_q$. 
% and $\delta$ is the Kronecker delta.
The updated posterior density (indicated by the prime on the updated parameters) for the SPAM fidelity ($f_q$)~\cite{gelman1995bayesian} :
\begin{equation}
f_q | \mathcal{D}_q \sim \text{Beta}(\alpha_q^\prime, \beta_q^\prime)
\end{equation}
where $\alpha_q^\prime = \alpha_q + L - C_0[\mathcal{D}_q]$ and $\beta_q^\prime = \beta_q + C_0[\mathcal{D}_q]$. This shows the influence of incremental measurements on posterior noise density. 
%, even with a single bit of information (L=1).
The qubit-wise updated mean~\cite{sivia2006data} of the SPAM fidelity, obtained as $\mu_q^\prime= \alpha_q^\prime / (\alpha_q^\prime+\beta_q^\prime)$, are then used in estimating the PEC coefficients.\\
%$v^\prime= \alpha_q^\prime\beta_q^\prime/ (\alpha_q^\prime+\beta_q^\prime)^2/(\alpha_q^\prime+\beta_q^\prime+1)$.
%This will be for the SPAM channel, which get used in the estimation of the quasi-probability distribution.\\
%
\subsection{Correlated parameters}
The second task involves the adaptive estimation of correlated noise parameters, which is more complex due to the measurements being influenced by multiple noise processes simultaneously. These are the noise parameters for which there exist datasets (typically the partial trace of an observable across a subset of qubits) which reflect errors from various noise parameters jointly. The analysis uses the probability of observing classical bit-strings on this qubit subset as the random variables.

For a subset of $m$ qubits from a total of $n$ qubits, there are $M = 2^m$ possible observed bit-strings, each with a probability denoted by $p_0, p_1, \ldots, p_{M-1}$, where $p_0$ represents the probability of observing all zeros and $p_{M-1}$ that of all ones. These probabilities are treated as random variables which form a probability simplex, as they are all positive and add up to 1. Hence the natural way to model the joint density is using a Dirichlet prior given by:
\begin{equation}
\text{Pr}(p_0, \cdots, p_{M-1}) = \frac{\Gamma(\sum a_i)}{\prod\limits_{i=0}^{M-1} \Gamma(a_i)} \prod\limits_{i=0}^{M-1} p_i^{a_i-1}
\end{equation}
where the gamma function $\Gamma(\cdot)$ was already defined previously and $\{a_i: a_i > 0, i \in \{0, \cdots M-1\}\}$ are the parameters characterizing the Dirichlet distribution that need to be estimated and updated -- a multi-variate task analogous to the uni-variate task for the beta distribution in the previous section.

We denote the dataset derived from $L$ measurements of the $m$ qubits in the computational basis as $\mathcal{D} = \{w(0), \cdots, w(L-1)\}$, where each $w(i)$ is a binary string of length $m$. 
The likelihood function is given by:
\begin{equation}
\mathcal{L}(p_0, \cdots, p_{M-1}) = 
\prod\limits_{i=0}^{M-1} p_i^{C_{v_i}[\mathcal{D}]}
\end{equation}
where $v_i \in \{0, 1\}^m$, 
$C_{v_i}[\mathcal{D}]$ is the count of occurrences of the string $v_i$ in the dataset $\mathcal{D}$. 
Note that $w(i)$ is a specific realization post-measurement that takes one of the values in the set $\{v_i\}$.

The posterior joint density is also a Dirichlet distribution~\cite{robert2010introducing}:
\begin{equation}
\Pr( p_0, \cdots, p_{M-1} \mid \mathcal{D}) = \frac{\Gamma(\sum a_i^\prime)}{\prod\limits_{i=0}^{M-1} \Gamma(a_i^\prime)} \prod\limits_{i=0}^{M-1} p_i^{a_i^\prime-1}
\label{eq:posterior_dirichlet}
\end{equation}
with parameters 
\begin{equation}
a_i^\prime = a_i + C_{v_i}(\mathcal{D}).
\end{equation}
Upon marginalizing this joint density, the marginals follow a Beta distribution. For example, the random variable $p_i$ is Beta-distributed with parameters~\cite{lee1989bayesian}: 
\begin{equation}
\begin{split}
\alpha_i =& a_i^\prime\\
\beta_i  =& -a_i^\prime+\sum\limits_{j=0}^{M-1} a_j^\prime
\end{split}
\end{equation}
The updated marginals from the Dirichlet distribution give us the evolving densities for each $p_i$, allowing us to calculate their time-varying means and variances. Specifically, the mean for $p_i$ updates to: 
\begin{equation}
\mathds{E}(p_i) = \frac{a_i^\prime}{\sum\limits_j a_j^\prime}
\end{equation}
and its variance updates to:
\begin{equation}
\text{Var}(p_i) = 
\frac{a_i^\prime\left(\sum\limits_j a_j^\prime - a_i^\prime\right)}
{\left(\sum\limits_j a_j^\prime\right)^2 \left(1+\sum\limits_j a_j^\prime\right)}.
\end{equation}
At the final stage, the method employs the relationship between the probabilities $p_0, \cdots, p_{M-1}$ and the noise model parameters to derive the time-dependent parameter means and variances of the noise parameters. 
% using $p_i = \Tr\left[\Pi_i 
%\Tr_{n-m} \left( \tilde{\mathcal{G}}_0 \rho  \right)
%\right]$, where $\Pi_i = \ket{v_i}\bra{v_i}$ is the projection operator corresponding to the bit-string $v_i$. The notation $\Tr_{n-m}(\cdot)$ represents a partial trace operation, where $n-m$ qubits have been traced out from the $n$-qubit density matrix, resulting in a reduced density operator for the $m$ qubits relevant to the correlated noise parameter estimation problem.
The process concludes by updating the quasi-probability distribution using the updated noise parameters per Eqn.~\ref{eq:basic_bayes}.
\subsection{Accuracy and stability}\label{sec:accuracy_and_stability}
We next evaluate the performance of PEC in presence of \nsn using the lens of accuracy and stability. A mean of a quantum observable $\mathcal{O}$ is represented as $\braket{\mathcal{O}}_\xrm$ (where $\xrm$ labels the noise instance), while the observed mean after mitigation is denoted as $\braket{\mathcal{O}}_\xrm^\text{mit}$, which equals $\braket{\mathcal{O}}$ only in the asymptotic limit of infinite samples and zero noise. To denote the mitigated noisy mean at a specific time $t$, we use $\braket{\mathcal{O}}_\xrm^\text{mit}(t)$.

We say that a PEC mitigated observable is $\epsilon-$accurate if the absolute difference between the mitigated and noiseless observable is upper bounded by $\epsilon$:
\begin{align}
&|\braket{O}_\xrm^\text{mit} - \braket{O}| \leq \epsilon
\end{align}

Now, suppose the underlying noise, being non-stationary, is characterized by a time-dependent density $f(\xrm;t)$. 
We say that a PEC mitigated observable is $\epsilon-$stable between times $t_1$ and $t_2$ if:
\begin{equation}
| \braket{O}_\xrm^\text{mit}(t_1) - \braket{O}_\xrm^\text{mit}(t_2) | \leq\epsilon
\label{eq:stability}
\end{equation}
where 
\begin{equation}
\begin{split}
&\braket{O}_\xrm^\text{mit}(t) = \int\limits_\xrm \braket{O}_\xrm^\text{mit} f(\xrm; t) d\xrm\\
&= \int\limits_\xrm 
\sum\limits_k 
\eta_\low{k}(\xrm, t)
\text{Tr} \left[ \mathcal{O} \tilde{\mathcal{G}}_k \rho \right] 
f(\xrm; t) d\xrm
\end{split}
\end{equation}
Note the dependence of $\eta_\low{k}$ on the random variable $\xrm$ and time $t$. 
When measuring $\mathcal{O}_q$ for qubit $q$, the qubit-wise accuracy metric is:
\begin{equation}
\epsilon_q = \left| \braket{\mathcal{O}_q}_\xrm - \braket{ \mathcal{O}_q } \right|
\end{equation}
without mitigation, and
\begin{equation}
\epsilon_q = \left| \braket{\mathcal{O}_q}_\xrm^\text{mit} - \braket{ \mathcal{O}_q } \right|
\end{equation}
with error mitigation. The register average, for a $n$-qubit register, is:
\begin{equation}
\epsilon_R = \sum\limits_{q=0}^{n-1} \epsilon_q / n
\label{eq:eR}
\end{equation}

Similarly, the qubit-wise stability metric is:
\begin{equation}
s_q(t) = \left| \braket{\mathcal{O}_q}_\xrm(t) - \braket{\mathcal{O}_q}_\xrm(0) \right|
\end{equation}
without mitigation, and
\begin{equation}
s_q(t) = \left| \braket{\mathcal{O}_q}_\xrm^\text{mit}(t) - \braket{\mathcal{O}_q}_\xrm^\text{mit}(0) \right|
\end{equation}
with error mitigation. The register average, for a $n$-qubit register, is:
\begin{equation}
s_R(t) = \sum\limits_{q=0}^{n-1} s_q(t)/n.
\label{eq:sR}
\end{equation}
We expect the accuracy ($\epsilon_q, \epsilon_R$) and stability $(s_q, s_R)$ metrics to be smaller when using adaptive PEC because of more accurate estimates for the time-varying PEC coefficients $\{ \eta_\low{k}\}$.
%
%\section{Analytical pec model under Pauli Channel assumption}
\section{Numerical Validation}\label{sec:PEC_under_Pauli}
%\section{Mitigating noise in \BV using adaptive PEC}\label{sec:PEC_under_Pauli}
%\subsection{Noiseless circuit}
%%%%%%%%%%%%%%%%%%%%%%%%%%%%%%%%%%%%%%%%%%%%%
\begin{figure}[htbp]
\centering
\includegraphics[width=\linewidth]{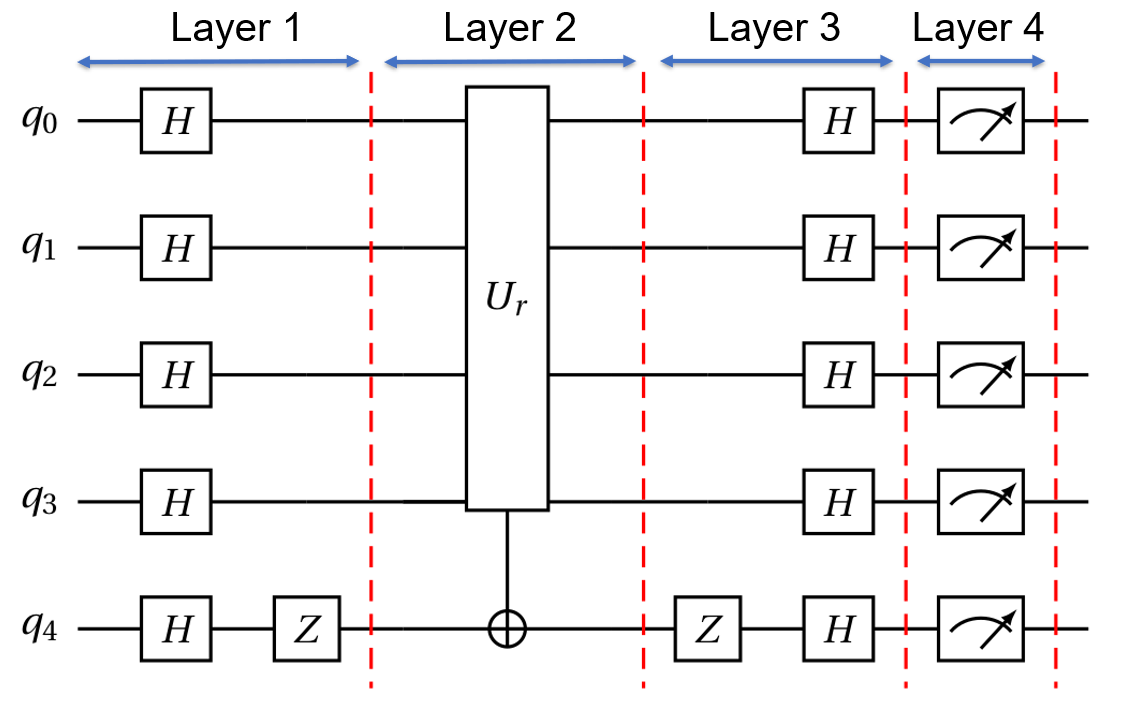}
\caption{A 5-qubit implementation of the Bernstein-Vazirani algorithm with secret bit-string $r$.}
%[Source code: myUtilities/qiskitUtils.py]
\label{fig:bv_circuit}
\end{figure}
%%%%%%%%%%%%%%%%%%%%%%%%%%%%%%%%%%%%%%%%%%%%%
For studying the stability and accuracy of PEC in presence of \nsn, we use an implementation of the Bernstein-Vazirani algorithm~\cite{bernstein1993quantum}, a standard benchmarking circuit that requires only a modest number of gates.
The purpose of the algorithm is to recover a $n$-bit secret string $r$, encoded in a black-box oracle function. 
The algorithm identifies the secret with a single query, while the classical method needs $2^n$ queries (worst-case). 

A quantum circuit for the \BV algorithm is shown in Fig.~\ref{fig:bv_circuit}. 
We conceptualize the circuit as having four principal layers ($\mathds{L}_1, \mathds{L}_2, \mathds{L}_3$, and $\mathds{L}_4$) separated by dashed vertical lines. 
%Two points to note with this circuit. First, the sequence of qubits $q_0 \cdots q_n$ is in reverse order compared to the binary representation $b_n \cdots b_0$ of the output string. 
Note that the total number of qubits needed by a $(n-1)$-bit secret string is $n$, comprising $n-1$ data qubits and one ancilla qubit. So, a 5-qubit implementation of the \BV algorithm has $n=5$ but the secret string length is $n-1=4$.

The circuit is initialized with all the qubits in the register in the $\ket{0}$ state. Thus, a pure state description yields an initial state $\ket{\psi_0}=\ket{0}^{\otimes n}$. 
In the first layer, all the qubits are subjected to Hadamard gates. The ancilla qubit (which is the last qubit with index n) is additionally subjected to a $\mathds{Z}$ gate. The output after the first layer $\mathds{L}_1 = \mathds{H}^{\otimes {n-1}} \otimes (\mathds{Z}_n\mathds{H}_n)$ is $\ket{\psi_1} = \ket{+}^{\otimes {n-1}}\ket{-}$. 

The second layer implements the oracle function for which we can use CNOT gates. 
Each CNOT's control qubit corresponds to one of the bits in the secret string $r$, while the target qubit remains fixed at qubit $n$. 
Specifically, if the bit $r_i$ of the secret string $r$ is 1, we add a CNOT between qubit $i$ (control) and qubit $n$ (target). 
The input to the second layer is $\mathds{H}^{\otimes {n-1}}\ket{0} \ket{-}$ while the output is $\ket{\psi_2} = \left( \mathds{H}^{\otimes {n-1}}\ket{r} \right)\ket{-}$ where $r$ is the secret string.

To retrieve the secret string $r$, the third layer requires another layer of Hadamard gates. 
A $\mathds{Z}$ gate is applied to the ancilla qubit (qubit $n$) before the application of the Hadamard layer to make the computing reversible. The output quantum state after the third layer $\mathds{L}_3 = \mathds{H}^{\otimes (n-1)} \otimes (\mathds{H}_n \mathds{Z}_n)$ is $\ket{\psi_3} = \ket{r}\ket{0}$.

\subsection{Observable}
The fourth layer is for measurement in the $\mathds{Z}$-basis and is not a unitary layer. We measure the state of each of the $n$-qubits after projection onto the computational basis states. 
Post-measurement, the observation obtained is a classical bit (either $0$ or $1$) for each of the n-qubits measured. 
Thus, the final observed output is a bit-string of length $n$, including the ancilla.

We make a distinction between the qubit-wise observables $\mathcal{O}_q$, where $q \in \{0, \cdots n\}$ and the measurement operator for the register  $M_Z = \mathds{Z}_0 \otimes \cdots \mathds{Z}_n$ in computational basis. 
The qubit-wise observable $\mathcal{O}_q$ has identity across the $n$-qubit tensor product except in the $q$-th position. 
For example, $\mathcal{O}_0 = \mathds{Z}_0 \otimes \mathds{I}^{\otimes {n-1}}$ and $\mathcal{O}_n = \mathds{I}^{\otimes {n-1}} \otimes \mathds{Z}_n$. 
The eigenvalues of $\mathcal{O}_q$ are +1 (corresponding to classical bit 0) and -1 (corresponding to classical bit 1). 
The theoretical mean of the qubit-wise observable $\mathcal{O}_q$ is denoted by $\braket{\mathcal{O}_q} = \Tr[\mathcal{O}_q \rho ]$ where $q \in \{0, \cdots, n \}$, which reflects the measurement process being fundamentally probabilistic. 

The experimentally observed mean of the observable, denoted by $\braket{\mathcal{O}_q}_\xrm$ is calculated as the sum of eigenvalues weighted by their empirically observed frequencies. In a noiseless circuit, the observed mean asymptotically converges to the theoretical mean as the sample size tends to infinity. However, in presence of shot noise and non-zero variance of $\xrm$, $\braket{\mathcal{O}_q}_\xrm$ may not equal $\braket{\mathcal{O}_q}$. 
%The subscript $\xrm$ denotes the presence of noise in general and reminds us of the empirical status of obtained mean of the observable.

%The simultaneous measurement of $\mathcal{O}_q$ across the qubits leads to the output quantum state being projected onto one of 
The experimentally observed measurements of the $M_Z$ operator in computational basis belong to one of the $2^{n}$ eigenstates, each of which can be represented by a $n$-bit binary string. These observations contain the information necessary for computing $\mathcal{O}_q$ for all $q$. 
In the context of the \BV algorithm, we discard the ancilla bit and declare the search a success when the first $n$-bits of the observed value of the $M_Z$ operator matches the secret string $r$. 
%However, this paper goes beyond just assessing the success rate of revealing $r$ through the initial $n$ bits; it focuses on evaluating the accuracy and stability of qubit-wise observables across the entire register, ancilla included.
%
\subsection{Modeling circuit noise}
Our experimental focus is on \superc~hardware in next section. The potential sources of noise in a \superc~implementation of the circuit depicted in Fig.~\ref{fig:bv_circuit} are: (i) state preparation noise, (ii) noise in the implementation of the Hadamard gate, (iii) noise in the implementation of the $\mathds{Z}$ gate, (iv) noise in the implementation of the CNOT gate, and (v) measurement noise (also known as readout noise). The error resulting from the first and last noise sources, namely state preparation noise and measurement noise, are often measured collectively due to experimental limitations. This combined noise is commonly referred to as SPAM (state preparation and measurement) noise. An effective model assumes that the state preparation is noiseless, and the noise impacts the measurement (or readout)  process. The relative magnitudes of the different types of noise are different. State preparation and measurement (SPAM) noise typically has the largest contribution to errors~\cite{bravyi2021mitigating}. The next most significant contribution arises from imperfect implementations of entangling gates, such as the CNOT. After that the strength of the noise, for single-qubit rotations, decreases by two orders of magnitude or more~\cite{van2023probabilistic, bravyi2021mitigating}. The $\mathds{Z}$ gate is a software-based operation~\cite{mckay2017efficient} and error-free. After disregarding single-qubit rotation errors, only two predominant types of noise emerge: SPAM noise and CNOT noise. 

%The single qubit rotation noise is of the relative order $10^{-4}$ compared to SPAM and CNOT noise.
%The justification for neglecting single-qubit errors is explained in Section~\ref{sec:single-qubit-errors}. 

Next, we create a quantum channel based description of the circuit noise. 
The circuit noise can be modeled layer-wise. 
The output density matrix prior to measurement can be represented as: 
$
\tilde{\rho} = 
\mathcal{E}_4( 
\mathcal{E}_3( 
\mathds{L}_3 \mathcal{E}_2(
\mathds{L}_2 \mathcal{E}_1(
\mathds{L}_1 \rho \mathds{L}_1^\dagger) 
\mathds{L}_2^\dagger) 
\mathds{L}_3^\dagger )
$
where $\mathcal{E}_k(\cdot)$ denotes the noise channel for the $k$-th layer of the circuit in Fig.~\ref{fig:bv_circuit}. We approximate $\mathcal{E}_1$ as identity channel because we ignore single-qubit errors. 

We model noise in the CNOT gate using a 2-qubit depolarizing model. 
Let $\xrm_C(t)$ and $\xrm_T(t)$ represent the stochastic depolarizing parameters for the control and target qubits at time t. 
The depolarizing noise model for the CNOT gate is represented by: 
\begin{equation}
\begin{split}
&\mathcal{E}_\text{CNOT}(\cdot) = [ 1-\xrm_T(t) ][ 1 - \xrm_C(t) ] (\cdot) +\\
&\frac{1-\xrm_C(t)}{3}\xrm_T \sum \limits_{\mathds{P}_T^\prime}  (\mathds{I}_C \otimes \mathds{P}_T^\prime) (\cdot) (\mathds{I}_C \otimes \mathds{P}_T^\prime)\\
&+\frac{1-\xrm_T(t)}{3}\xrm_C \sum \limits_{\mathds{P}_C^\prime}  (\mathds{P}_C^\prime \otimes I_T ) (\cdot) (\mathds{P}_C^\prime \otimes I_T )\\
&+\frac{\xrm_C(t) \xrm_T }{9} \sum \limits_{\mathds{P}_C^\prime, \mathds{P}_T^\prime}  (\mathds{P}_C^\prime \otimes \mathds{P}_T^\prime) (\cdot) (\mathds{P}_C^\prime \otimes \mathds{P}_T^\prime)
\end{split}
\end{equation}
In this expression, we use $\mathds{P}_C^\prime, \mathds{P}_T^\prime \in \{\mathds{X}, \mathds{Y}, \mathds{Z}\}$ as the single-qubit Pauli operators excluding identity $\mathds{I}$, acting on the control (C) and target (T) qubits respectively. The sum is over all the single-qubit Pauli operators, excluding identity. Note that the effect of the quantum noise channel $\mathcal{E}_2(\cdot)$ for layer 2, on a 5-qubit state, combines the identity channel $\mathcal{I}$ (which acts on qubits without CNOT connection) and $\mathcal{E}_\text{CNOT}$ (which acts on qubits linked by CNOT connections). 

Similar to layer 1, layer 3 also comprises single-qubit rotations only. Hence, we treat $\mathcal{E}_3(\cdot)$ as identity channel. Lastly, the fourth layer is subjected to SPAM noise which we adapt to a noise channel description~\cite{smith2021qubit}. \added{
It effectively handles measurement noise by corrupting the density matrix post execution but pre measurement, and then conducting noise-free projective measurements on the corrupted output. }
The SPAM noise channel for qubit $q$ has two Kraus operators $M_0$ and $M_1$:
\begin{equation}
\begin{split}
M_0 =& \sqrt{ f_q}\ket{0}\bra{0}+\sqrt{1-f_q}\ket{1}\bra{1}\\
M_1 =& \sqrt{1-f_q}\ket{0}\bra{0}+\sqrt{f_q}\ket{1}\bra{1}\\
\end{split}
\end{equation}
where $f_q$ represents the SPAM fidelity of qubit $q$. 
% which is the probability of successfully measuring qubit $q$ when prepared in a fiducial state. 
The probability of observing $0$ is given by $\text{Tr}[M_0^\dagger M_0 \rho]$ and the probability of observing $1$ is given by $\text{Tr}[M_1^\dagger M_1 \rho]$. Note that we have assumed a symmetric model for SPAM noise with $f_q$ denoting the average SPAM fidelities for the initial states prepared as $\ket{0}$ and $\ket{1}$.

%Suppose the SPAM fidelity for each of the $n+1$-qubits in the register is denoted by $f_0, \cdots f_n$. 
Neglecting inter-qubit cross-talk, we then have the noise channel representation for the last layer $\mathcal{E}_4$ as a separable SPAM noise channel:
\begin{equation}
\mathcal{E}_4(\cdot) = \left[\bigotimes\limits_{q=0}^{n} \mathcal{E}_q^\text{SPAM}\right](\cdot)
\end{equation}
%The way to intepret this expression is as follows. 
%When provided with the density matrix representing a quantum state of $n$-qubits, the operation $\mathcal{E}_4$ has the following effect: the q-th qubit is affected by a single-qubit SPAM noise channel, with Kraus operators ($M_0$ and $M_1$) defined by the SPAM fidelity $f_q$. 
%
\subsection{PEC coefficients under Pauli noise}\label{sec:analytical_BV_model}
We first discuss CNOT noise mitigation using PEC. 
Following that, we discuss single-qubit SPAM noise mitigation using PEC. 
Then, we integrate the two discussions for noise mitigation in the quantum circuit implementation of the \BV algorithm using PEC. 
%Throughout this discussion, we employ the Pauli channel as an effective noise model.\\
%
\subsubsection{CNOT noise}\label{sec:cnotPEC}
Let $\tilde{\mathcal{G}_0}$ denote the \sop~for the noisy CNOT operation. 
Under the Pauli channel assumption, there are 16 basis {\sop}s, indexed by $k$:
\begin{equation}
\tilde{\mathcal{G}_k} \rho = (\mathds{P}_C \otimes \mathds{P}_T) \left( \tilde{\mathcal{G}_0} \rho \right) (\mathds{P}_C \otimes \mathds{P}_T)
\end{equation}
where $\rho$ is a 2-qubit density matrix, 
$\mathds{P}_C, \mathds{P}_T \in \{ \mathds{I},\mathds{X},\mathds{Y},\mathds{Z}\}$ are the single-qubit Paulis acting on the control (C) and target (T) qubits respectively, 
and $k \in \{0, \cdots 15\}$. 
The index $k$ is determined by the specific combination of $\mathds{P}_C$ and $\mathds{P}_T$, starting from $\mathds{I}\otimes\mathds{I}$ for $k=0$ and ending with $\mathds{Z}\otimes\mathds{Z}$ for $k=15$, incrementing $k$ for each subsequent combination in the sequence $\mathds{I}, \mathds{X}, \mathds{Y}, \mathds{Z}$ applied to $\mathds{P}_C$ and $\mathds{P}_T$. 

\begin{figure}[htbp]
\centering
\includegraphics[width=.2\linewidth]{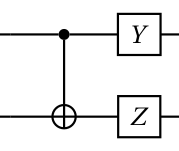}
\caption{Circuit diagram for implementing one of the 16 noisy {\sop}s for CNOT noise mitigation using PEC. The initial noisy CNOT gate is followed up by $\mathds{Y}$ and $\mathds{Z}$ gates on the control and target qubits respectively.}
\label{fig:CNOT_YZ}
\end{figure}

When $\mathds{P}_C = \mathds{P}_T = \mathds{I}$, the circuit corresponds to the noisy \sop~$\tilde{\mathcal{G}_0}$ denoting the noisy CNOT gate available. 
An example of one of the remaining 15 PEC circuits for CNOT noise mitigation is shown in Fig.~\ref{fig:CNOT_YZ}.
%%%%%%%%%%%%%%%%%%%%%%%%%%%%%%%%%%%%%%%%%%%%%%%%%%%%%%%%%%%%%%%%%%%%%%%%
%%%%%%%%%%%%%%%%%%%%%%%%%%%%%%%%%%%%%%%%%%%%%%%%%%%%%%%%%%%%%%%%%%%%%%%%

%\mathds{U}_\text{CNOT} \rho \mathds{U}_\text{CNOT}^\dagger
Using the requirement that the linear combination $\sum \eta_{ {}_k} \tilde{\mathcal{G}}_k $ should equal the noiseless CNOT operation, we derive the PEC coefficients as follows. When $\mathds{P}_C = \mathds{P}_T = \mathds{I}$, $\eta_\low{0}  = c_0c_1$. When $\mathds{P}_C=\mathds{I}$ and $\mathds{P}_T \in \{\mathds{X}, \mathds{Y}, \mathds{Z}\}$, $\eta_\low{k} = c_0 \xrm_T (1-\xrm_C)/3, \;\;\;\; k \in \{1,2,3\}$. When $\mathds{P}_T=\mathds{I}$ and $\mathds{P}_C \in \{\mathds{X}, \mathds{Y}, \mathds{Z}\}$, $\eta_\low{k} = c_0 \xrm_C (1-\xrm_T) /3, \;\;\;\; k \in \{4,8,12\}$. For all the remaining terms, $\eta_\low{k} = c_0 \xrm_C \xrm_T/9, \;\;\;\; k \in \{9, 10, 11, 13, 14, 15\}$. Here, $c_0 = 3/[\xrm_T (1-\xrm_C)]^2+[\xrm_C (1-\xrm_T)]^2-3c_1^2]$ and, $c_1 = \xrm_C + \xrm_T - \xrm_C \xrm_T - 1$. 
%[source code: myUtilities/sympyUtils/getEqns4CNOTpec()]. 
This yields the quasi-probability distribution as: 
\begin{equation}
\{\mathds{Q}_k\} = \left\{ |\eta_{{}_0}|/\gamma_{{}_\text{CNOT}}, \cdots, |\eta_{{}_{15}}|/\gamma_{{}_\text{CNOT}} \right\}
\end{equation}
where $\gamma_{{}_\text{CNOT}} = |\eta_{{}_0}| + \cdots + |\eta_{{}_{15}}|.$\\
\subsubsection{SPAM noise for qubit q}\label{sec:SPAMpec}
Consider a single-qubit SPAM noise channel for qubit $q$. The four noisy {\sop}s that can be implemented in this case are: 
(i) the SPAM noise channel: $\tilde{\mathcal{G}}_\mathds{I}(\rho) = f_q\rho + (1-f_q)\mathds{X}\rho \mathds{X}$, 
(ii) the SPAM noise channel followed by an X error: $\tilde{\mathcal{G}}_\mathds{X}(\rho) = f_q\mathds{X}\rho \mathds{X} + (1-f_q)\rho$, 
(iii) the SPAM noise channel followed by a Y error: $\tilde{\mathcal{G}}_\mathds{Y}(\rho) = f_q\mathds{Y}\rho \mathds{Y} + (1-f_q)\mathds{Z}\rho \mathds{Z}$ and, 
(iv) the SPAM noise channel followed by a Z error: $\tilde{\mathcal{G}}_\mathds{Z}(\rho) = f_q\mathds{Z}\rho \mathds{Z} + (1-f_q)\mathds{Y}\rho \mathds{Y}$.

Solving the linear equation: 
\begin{equation}
\mathcal{I} = \eta_0 \tilde{\mathcal{G}}_I + \eta_1 \tilde{\mathcal{G}}_X + \eta_2 \tilde{\mathcal{G}}_Y + \eta_3 \tilde{\mathcal{G}}_Z
\label{eq:pec_eq_spam}
\end{equation}
we get the quasi-probability distribution as:
$
\left\{ |\eta_\low{0}|/\gamma_{{}_\text{SPAM}}, |\eta_\low{1}|/\gamma_{{}_\text{SPAM}}, 0, 0 \right\}
$
with
\begin{equation}
\begin{split}
&\eta_\low{0} = \frac{f_q}{2f_q-1}, \;\;\;\; \eta_\low{1} = -\frac{1-f_q}{2f_q-1}, \;\;\;\; \eta_\low{2} = \eta_\low{3} = 0\\
&\text{sgn}(\eta_\low{0}) = +1, \;\;\;\; \text{sgn}(\eta_\low{1}) =-1\\
&\gamma_{{}_\text{SPAM}}(q) = |\eta_\low{0}|+|\eta_\low{1}|\\
\end{split}
\label{eq:SPAM_PEC}
\end{equation}
Estimating $f_q$ therefore provides a complete specification of the PEC coefficients.\\
\subsubsection{Composite noise in 5-qubit implementation of \BV algorithm}
The two noisy basis {\sop}s for each of the 5 distinct SPAM noise channels and the 16 noisy basis {\sop}s for the CNOT noise channel leads to 512 ($16\times 2^5$) noisy basis circuits $\{ \mathcal{G}_k \}$ for the 5-qubit \BV circuit, where $k$ runs from 0 to 511. 
%One of these noisy basis circuits is shown in Fig.~\ref{fig:bv_circuit_XIXXIYZ}.
%%%%%%%%%%%%%%%%%%%%%%%%%%%%%%%%%%%%%%%%%%%%%%%%%%%%%%%%%%%%%%%%%%%%%%%%%
%\begin{figure}[htbp]
%\centering
%\includegraphics[width=\linewidth]{bv_circuit_XIXXIYZ.png}
%\caption{One of the 512 noisy basis circuits for mitigation using PEC in the 5-qubit implementation of the \BV circuit. The initial noisy CNOT gate is followed up with $\mathds{Y}$ and $\mathds{Z}$ gates on the control and target qubits respectively. The readout lines for qubit $0, 1, 2, 3, 4$, and $5$ are subjected to the Pauli gates $\mathds{X}, \mathds{I}, \mathds{X}, \mathds{X}, \mathds{I}$ respetively, prior to measurement, in this specific noisy basis circuit for PEC.}
%\label{fig:bv_circuit_XIXXIYZ}
%\end{figure}
%%%%%%%%%%%%%%%%%%%%%%%%%%%%%%%%%%%%%%%%%%%%%%%%%%%%%%%%%%%%%%%%%%%%%%%%%

%We denote the mean of the quantum observable $\mathcal{O}$ obtained from the k-th noisy basis circuit by:
%$\braket{\mathcal{O}_{k}}_\xrm = \text{Tr}\left[
%\mathcal{O} \tilde{\mathcal{G}}^\text{BV}_k
%\rho_0\right]$, where $\rho_0$ is the input to the circuit represented by the {\sop} $\mathcal{G}^\text{BV}_k$ and $\xrm$ denotes presence of noise.

%These means are linearly combined to estimate the mean of the observable for the noiseless circuit:
%\begin{equation}
%\braket{\mathcal{O}}^\text{mit}_\xrm = \gamma 
%\sum \limits_{k}
%\mathds{Q}_\lowk
%\text{sgn}\left( 
%\eta^\text{BV}_\lowk
%\right)
%\braket{\mathcal{O}_\lowk}_\xrm
%\end{equation}
Under the SPAM noise separability assumption, $\gamma$ is obtained as:
\begin{equation}
\gamma = \gamma_{{}_{CNOT}} \prod\limits_{q=0}^{4} \gamma_{{}_{SPAM}} (q)
\end{equation}
where $\gamma_{{}_{SPAM}} (q)$ refers to the $\gamma_{{}_{SPAM}}$ for the $q$-th qubit. The PEC coefficients for the \BV circuit are the elements of:
\begin{equation}
%\left\{ \eta_k^\text{BV} \right\} = 
\left\{ \eta_\low{0}^\text{CNOT}, \cdots, \eta_\low{15}^\text{CNOT} \right\} 
\times \prod\limits_{q=0}^{4}
\left\{ \eta_\low{0}^\text{SPAM}(q),  \eta_\low{1}^\text{SPAM}(q) \right\}
\label{eq:cartesian_product}
\end{equation}
The quasi-probability distribution then follows as: $\left\{\mathds{Q}_k\right\} = \left\{|\eta_\low{k}^\text{BV}| /\gamma \right\}$.\\

%where $\gamma = \sum\limits_k |\eta_\low{k}^\text{BV}|$. 
%
%%%%%%%%%%%%%%%%%%%%%%%%%%%%%%%%%%%%%%%%%%%%
\begin{figure*}
\centering
\begin{minipage}{0.49\textwidth}
  \centering
  \includegraphics[width=\linewidth]{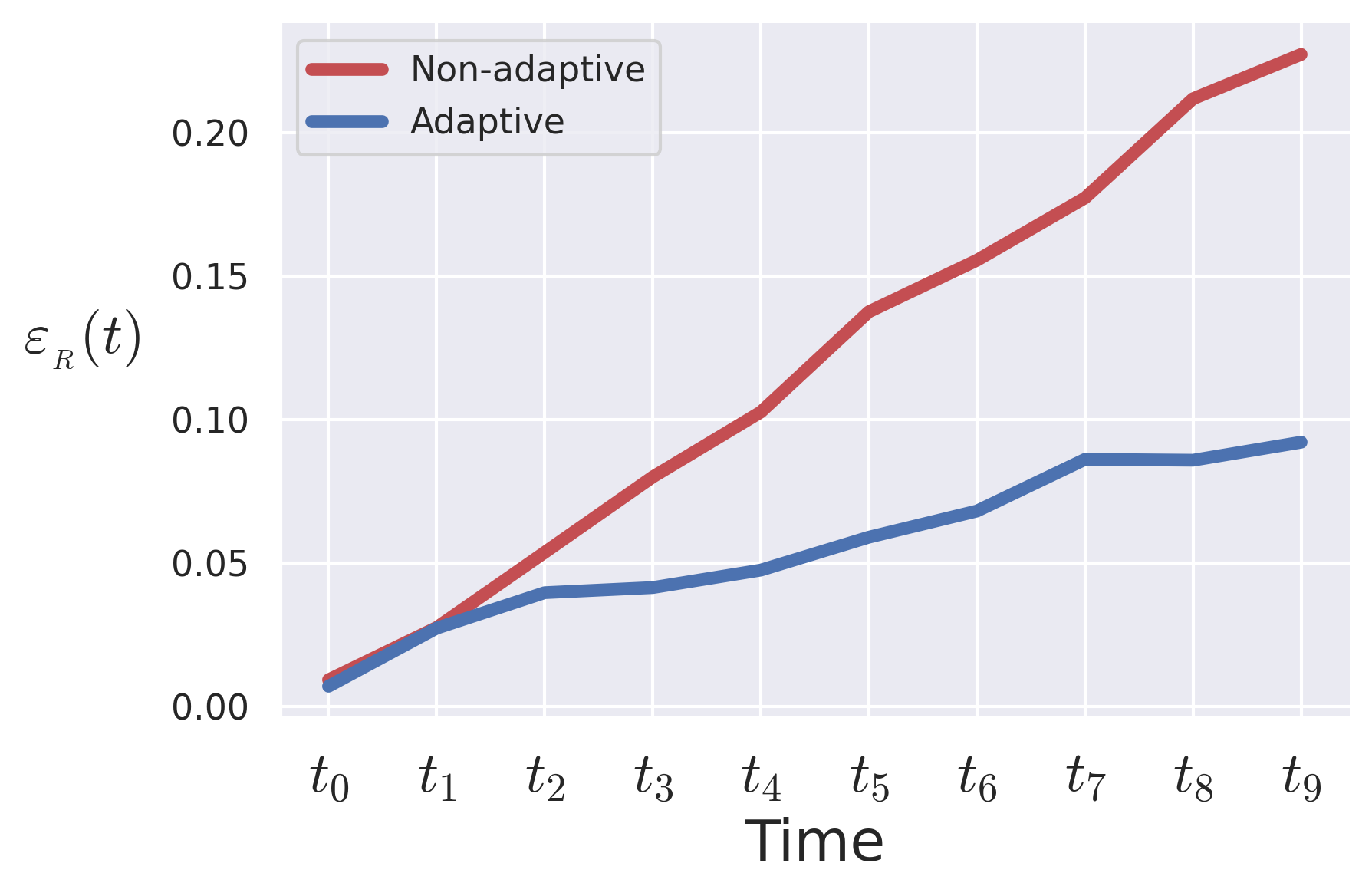}
  \caption*{(a)}
\end{minipage}
\hfill
\begin{minipage}{0.49\textwidth}
  \centering
  \includegraphics[width=\linewidth]{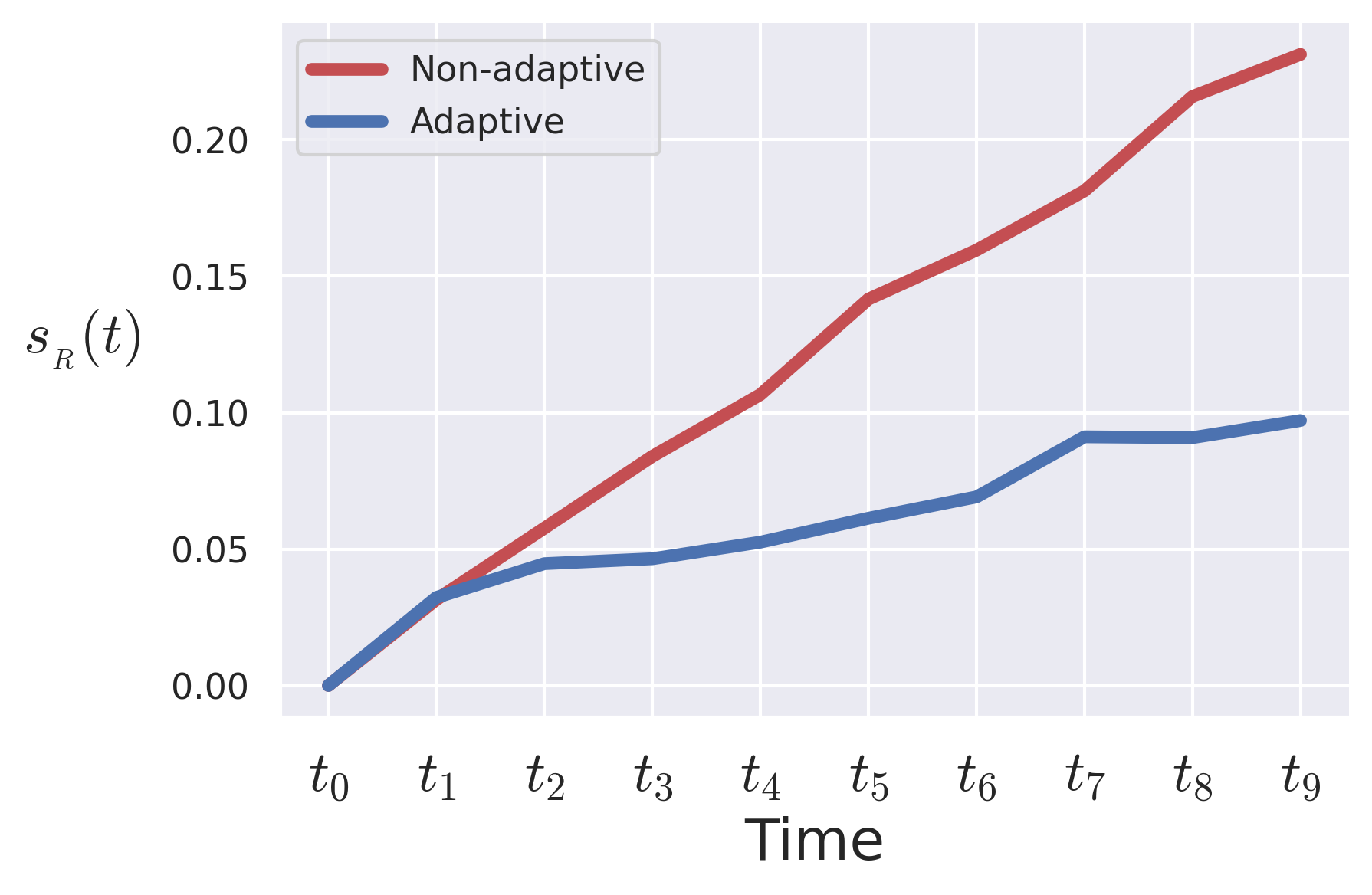}
  \caption*{(b)}
\end{minipage}
\caption{
Comparison of the adaptive PEC algorithm with a non-adaptive approach for a density matrix simulation that implements a noisy 5-qubit circuit for solving the Bernstein Vazirani problem. (a) Average accuracy across the 5-qubits and (b) Average stability across the 5-qubits.
% [Source code: myUtils/pecUtils/pecInstability()]
}
\label{fig:num_sim_stability_accuracy}
\end{figure*}
%\FloatBarrier
%%%%%%%%%%%%%%%%%%%%%%%%%%%%%%%%%%%%%%%%%%%%
\subsection{Adaptive noise model}\label{sec:bayesian-stabilization}
In the context of our 5-qubit Bernstein-Vazirani setup, the noise parameters $f_0$, $f_1$, and $f_2$ are estimated using the method for uncorrelated parameters as they do not have CNOT correlations. The adaptive estimation of parameters $f_3, f_4, \xrm_C,$ and $\xrm_T$ employs the method for handling correlated parameters, detailed in Sec.~\ref{sec:adaptPEC}. 
The process is initiated within the Bayesian inference framework, which considers the probabilities of observing outcomes 00, 01, 10, and 11 on qubits 3 and 4, as depicted in Fig.~\ref{fig:bv_circuit}, to be random variables. 
The estimation of the time-varying means and variances of these correlated noise parameters ($f_3, f_4, \xrm_C, \xrm_T$) is achieved by associating the mean values of the estimated densities directly with the parameters of the correlated noise model using:
\begin{equation}
\Pr(i) = \Tr\left[
\Pi_i
\mathcal{E}_4( 
\mathds{L}_3 \mathcal{E}_2
(\mathds{L}_2 \mathds{L}_1 \rho \mathds{L}_1^\dagger \mathds{L}_2^\dagger) 
\mathds{L}_3^\dagger
)
\right]
\end{equation}
where $\Pi_i = \ket{i}\bra{i}$ are the projection operators and $i \in \{00, 01, 10, 11\}$.
For our running example, the probabilities are given by:
\begin{equation}
\begin{split}
\text{Pr}(00) =& f_3 f_4 (-1 - \xrm_C \xrm_T + \xrm_C+\xrm_T)\\
&+ f_3 (\xrm_C \xrm_T /2-\xrm_T/2) \\
&+ f_4 (1 -\xrm_T - \xrm_C /2  + \xrm_C \xrm_T /2)\\
&- \xrm_C \xrm_T/4 +\xrm_T/2\\
\text{Pr}(01) = & f_3 f_4 (1 -\xrm_C -\xrm_T + \xrm_C \xrm_T)\\
&+ f_3 (-1 -\xrm_C \xrm_T /2 + \xrm_C   - \xrm_T /2)\\
&+ f_4 (-1 -\xrm_C \xrm_T /2 + \xrm_C   +\xrm_T)\\
&+1 +\xrm_C \xrm_T /4  -\xrm_C/2 -\xrm_T /2\\
\text{Pr}(10) = & f_3 f_4 (1-\xrm_T -\xrm_C +\xrm_C \xrm_T )\\
&+f_3 (\xrm_T /2-\xrm_C \xrm_T /2)\\
&+f_4 (\xrm_C /2 -\xrm_C \xrm_T /2)\\
&+ \xrm_C \xrm_T /4\\
\text{Pr}(11) = & f_3 f_4 (-1 +\xrm_C +\xrm_T - \xrm_C\xrm_T)\\
&+ f_3 (1 -\xrm_T/2 -\xrm_C +\xrm_C\xrm_T /2)\\
&+ f_4 (\xrm_C\xrm_T/2 -\xrm_C/2)\\
&+\xrm_C/2 - \xrm_C\xrm_T /4\\
\label{eq:tv-correlated-means}
\end{split}
\end{equation}
%[Source code: myUtils/sympyUtils/getNoisyCNOToutput()]

In the last step, the updated PEC coefficients are obtained using Eqn.~\ref{eq:cartesian_product}. 
\subsection{Numerical simulation}
%Recall that the qubit-wise mean of the observable obtained post-mitigation is denoted by:
%\begin{equation}
%\braket{\mathcal{O}_q}_\xrm^\text{mit} = \gamma \mathds{E} \left[
%\text{sgn}(\eta^\text{BV}_\low{k}) \text{Tr}
%\left[
%\mathcal{O}_q \tilde{G}^\text{BV}_k(\rho)
%\right]
%\right]
%\end{equation}
%where 
%$q \in \{0, \cdots n-1 \}$ and $n$ is the size of the quantum register. 
For our 5-qubit circuit implementing the \BV algorithm, we used secret string $r=1000$. 
Thus, the qubit-wise mean of the $\mathds{Z}$ observable for the noiseless case is given by: $\braket{ \mathcal{O}_0 } = +1, \braket{ \mathcal{O}_1 } = +1, \braket{ \mathcal{O}_2 } = +1, \braket{ \mathcal{O}_3 } = -1, \braket{ \mathcal{O}_4 } = +1$, qubit 4 being the ancilla. 

To validate our method, we used a numerical experiment that conducts a density matrix simulation of a 5-qubit noisy quantum circuit implementing the Bernstein-Vazirani algorithm using the Qiskit~\cite{alexander2020qiskit} software. 
The simulation begins with the mean of the beta distributions characterizing the SPAM fidelities for qubits 0-4 set at $0.96, 0.95, 0.94, 0.93, 0.92$, respectively, and the mean of the depolarizing channel parameters for the control and target qubits in the CNOT gate both fixed at $0.017$. Over the course of ten simulated time periods, the average SPAM fidelity for each qubit decreased by 0.01 per period, resulting in final mean SPAM fidelities of $0.86, 0.85, 0.84, 0.83, 0.82$ for qubits 0-4, respectively. Similarly, the average depolarizing parameter for the CNOT gate also declined by 0.01 per time period, leading to a final mean value of $0.117$ for both control and target qubits by the simulation's end. The noise parameters were adaptively estimated, per the methodology described in previous sections, using data generated by executing each of the 512 PEC circuits using $10,000$ shots.

Fig.~\ref{fig:num_sim_stability_accuracy} demonstrates the simulated efficacy of the adaptive PEC algorithm. The plot compares the register accuracy and stability (averaged over the 5 qubits) achieved with the adaptive approach against a non-adaptive approach. It shows an improvement in accuracy of 59.5\% in the final time period, when device noise is at its peak, with an average accuracy improvement of 53.4\% across all ten periods. Similarly, stability improved by 58.0\% in the last period, and by an average of 51.5\% over the entire span of ten periods. The improvement in accuracy and stability of the outcomes from adaptive PEC occur due to more accurate noise characterizations using Bayesian inference.
\section{Experimental Testing}\label{sec:real_data}
%%%%%%%%%%%%%%%%%%%%%%%%%%%%%%%%%%%%%%%%%%%%%%%%%%%%%%%%%%%%%%%%%%%%%%%%
\begin{figure}[htbp]
\centering
\includegraphics[width=\linewidth]{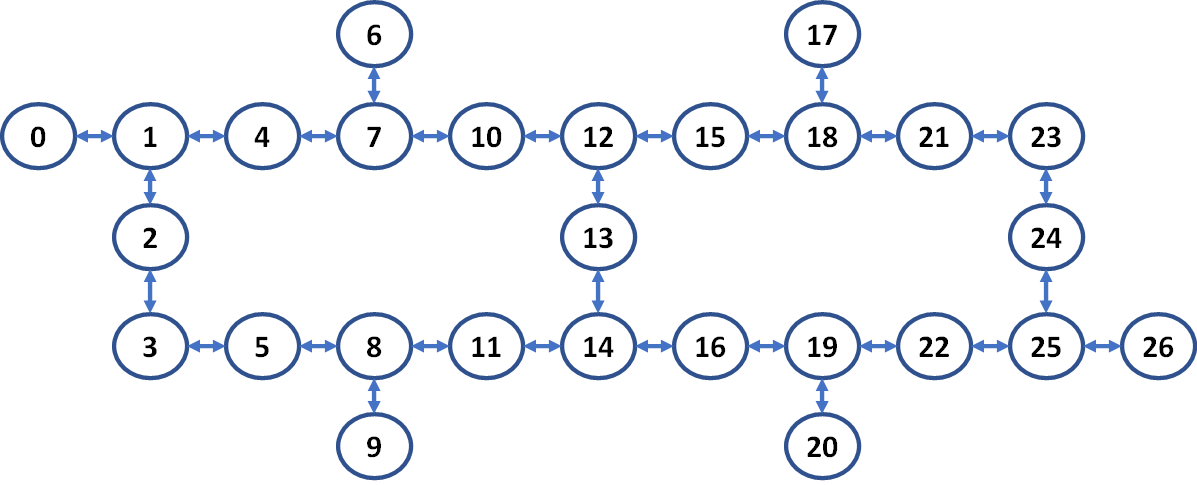}
\caption{Qubit layout of the 27-qubit \superc~device \kolkata. Circles represent superconducting transmons and lines indicate possible gate operations between sites.}
\label{fig:kolkata4}
\end{figure}
%%%%%%%%%%%%%%%%%%%%%%%%%%%%%%%%%%%%%%%%%%%%%%%%%%%%%%%%%%%%%%%%%%%%%%%%
%%%%%%%%%%%%%%%%%%%%%%%%%%%%%%%%%%%%%%%%%%%%%
\begin{figure*}
\centering
\begin{minipage}{0.8\textwidth}
  \centering
  \includegraphics[width=\linewidth]{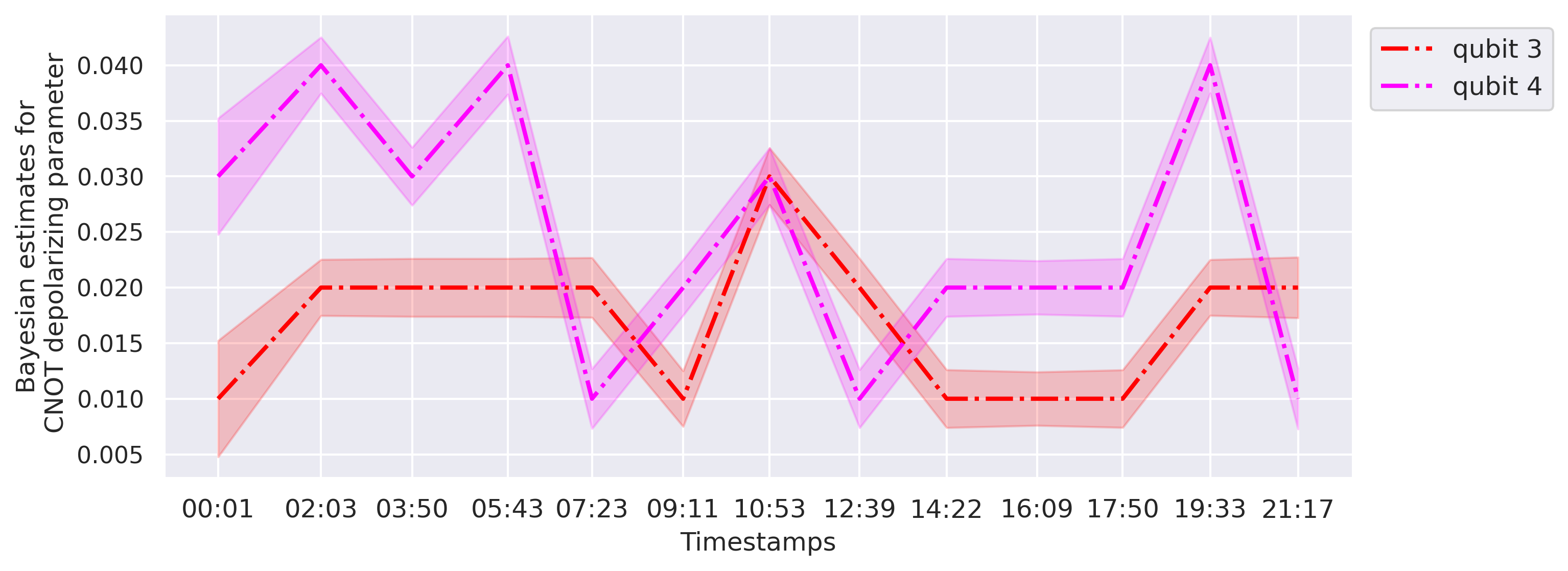}
  \caption*{(a)}
\end{minipage}
\hfill
\begin{minipage}{0.8\textwidth}
  \centering
  \includegraphics[width=\linewidth]{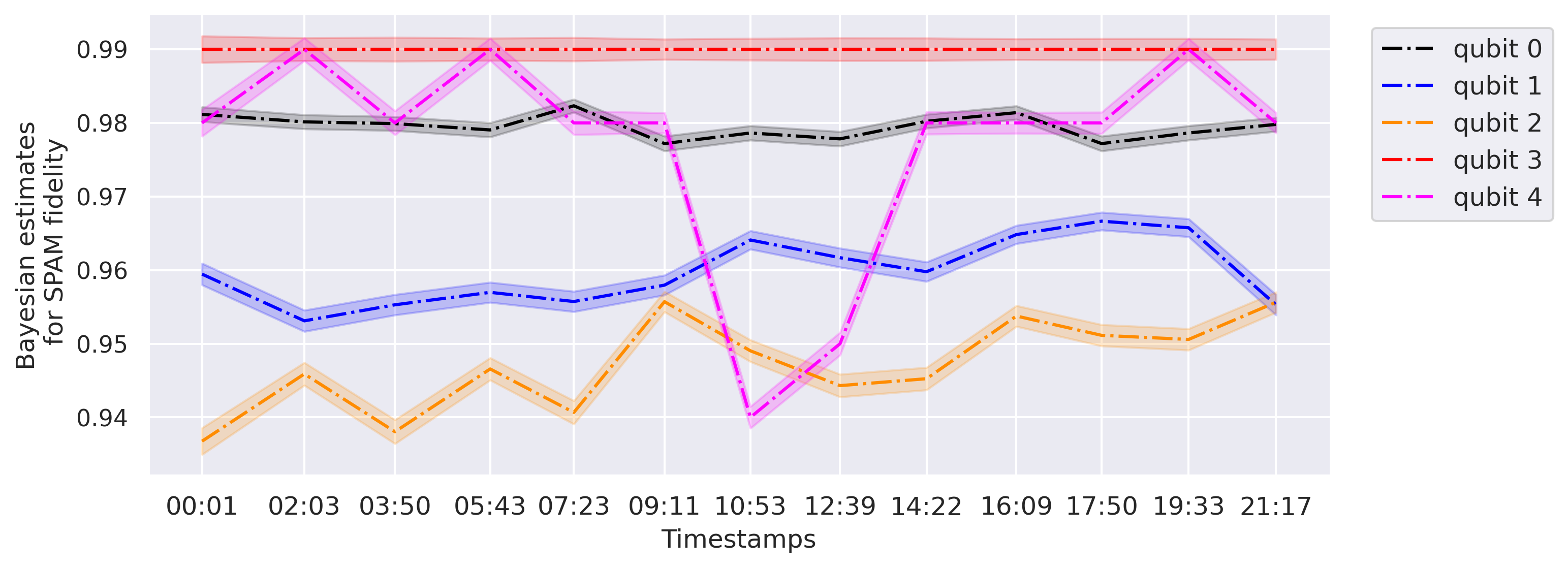}
  \caption*{(b)}
\end{minipage}
\caption{
This figure depicts the \nsn on the experimental device \kolkata. In plot (a), the blue line represents the depolarizing parameter for the target qubit, while the black line denotes the depolarizing parameter for the control qubit in the CNOT gate. Plot (b) illustrates five lines, each indicating the SPAM fidelity for the register elements. The x-axis corresponds to intra-calibration timestamps for January 15. The shaded regions denote the time-varying standard deviations.
}
\label{fig:tv_bv_depol_spam}
\end{figure*}
%\FloatBarrier
%%%%%%%%%%%%%%%%%%%%%%%%%%%%%%%%%%%%%%%%%%%%%
We tested the adaptive PEC method on the 27-qubit \superc~device called \kolkata.  
Qubits 0,1,2,3,4 in Fig.~\ref{fig:bv_circuit} map to physical qubits 0,1,2,3,5 on the device shown in Fig.~\ref{fig:kolkata4}. 
The CNOT gate is between the physical qubits 3 (control) and 5 (target) on \kolkata. 
Our dataset spans 24 hours and comprises 13 complete PEC datasets. 
It was collected on January 15, 2024, and have the following time-stamps: 00:01 hrs, 02:03 hrs, 03:50 hrs, 05:43 hrs, 07:23 hrs, 09:11 hrs, 10:53 hrs, 12:39 hrs, 14:22 hrs, 16:09 hrs, 17:50 hrs, 19:33 hrs, and 21:17 hrs. 
Each dataset is derived from measurements made in the computational basis, with each observation being a 5-bit string. 
The observations were obtained from the 512 noisy basis circuits, 
as described in Sec.~\ref{sec:analytical_BV_model}, 
with each circuit repeated using $L=10,000$ shots, 
resulting in approximately 67 million observations in total.
%
%%%%%%%%%%%%%%%%%%%%%%%%%%%%%%%%%%%%%%%%%%%%%
\begin{figure*}
\centering
\begin{minipage}{0.8\textwidth}
  \centering
  \includegraphics[width=\linewidth]{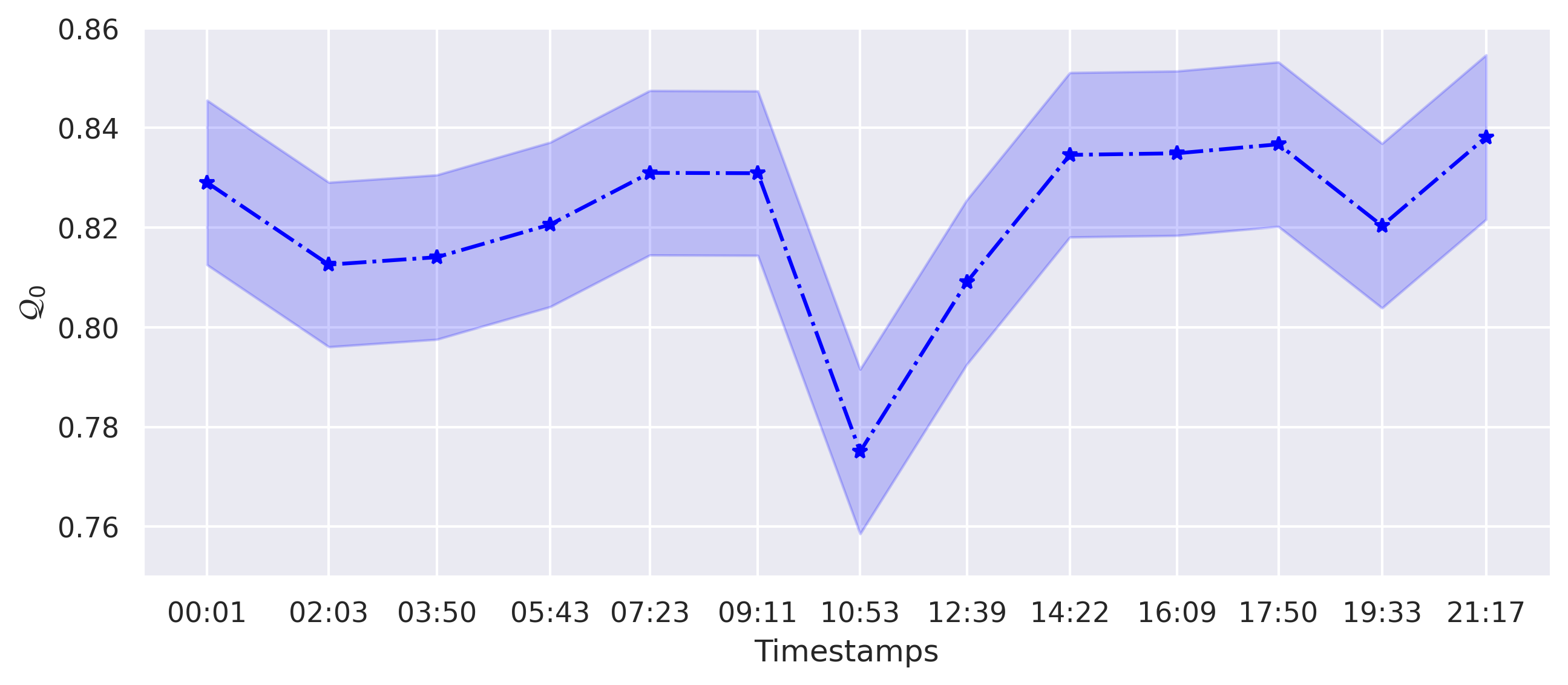}
  \caption*{(a)}
\end{minipage}
\hfill
\begin{minipage}{0.8\textwidth}
  \centering
  \includegraphics[width=\linewidth]{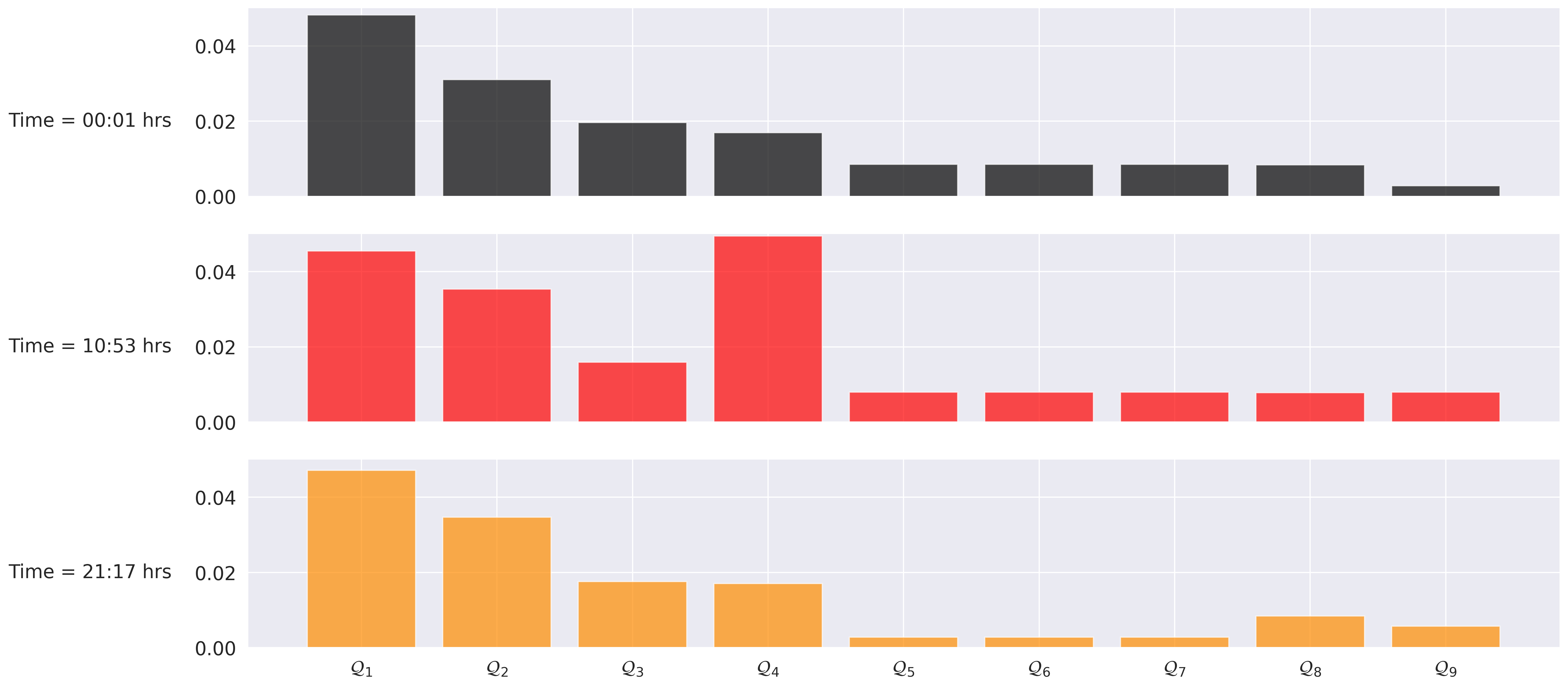}
  \caption*{(b)}
\end{minipage}
\caption{
These plots depict the \ns~nature of the quasi-probability distribution. To maintain clarity, we opted not to plot all 512 bins of the histogram in a single plot. In (a), we observe the time-varying weight $\mathds{Q}_0$ for the raw Bernstein-Vazirani circuit which carries the most substantial weight, accounting for almost 80\% of the distribution. In (b), we show the values of the quasi-probability bins for the subsequent 10 basis circuits, which are more than 10 times lower in magnitude compared to the first circuit. The \ns~nature of the quasi-probability distribution becomes crucial given the lengthy data collection process required for PEC mitigation because the noise estimation becomes inaccurate in these time-frames. Our experiment took approximately 2 hours for each dataset comprising 512 circuits.
}
\label{fig:tv_qpr}
\end{figure*}
%\FloatBarrier
%%%%%%%%%%%%%%%%%%%%%%%%%%%%%%%%%%%%%%%%%%%%%
%%%%%%%%%%%%%%%%%%%%%%%%%%%%%%%%%%%%%%%%%%%%%
\begin{figure*}
\centering
\begin{minipage}{.7\textwidth}
  \centering
  \includegraphics[width=\linewidth]{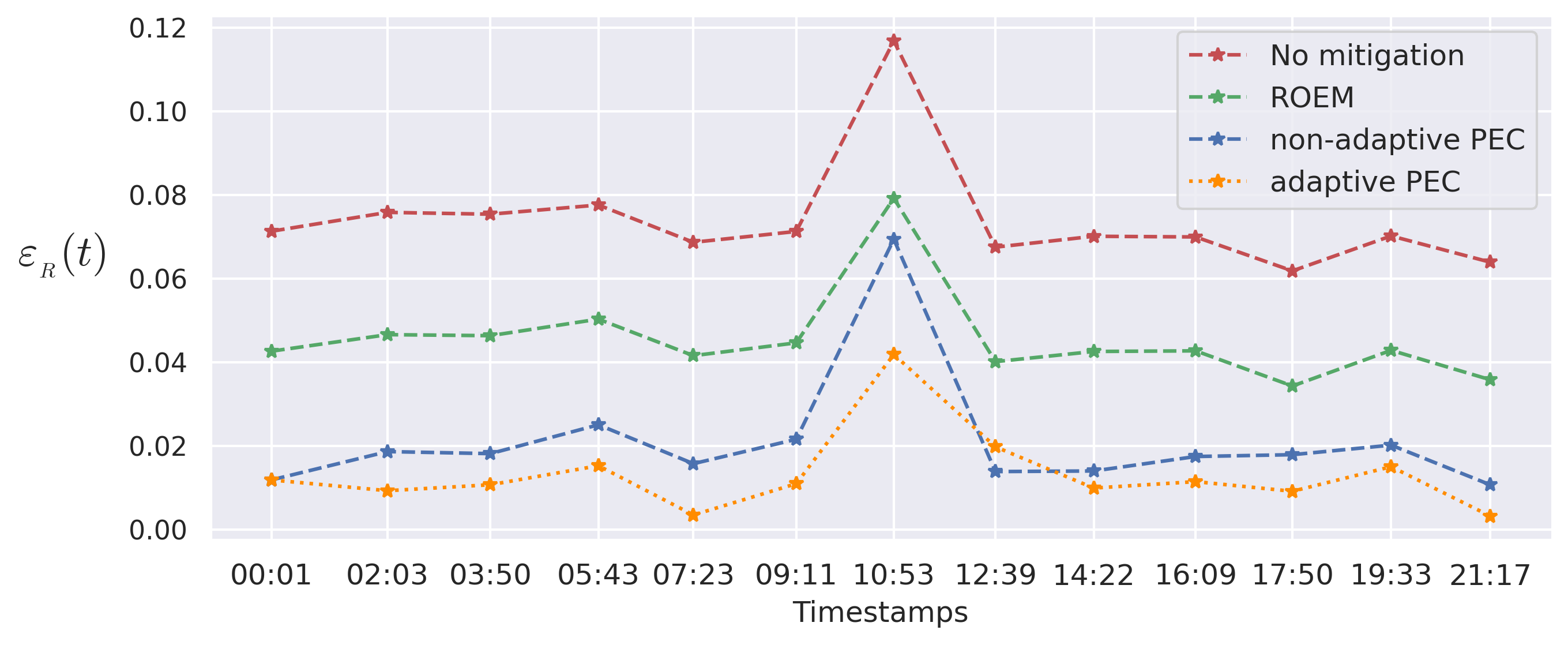}
  \caption*{(a)}
\end{minipage}
\hfill
\begin{minipage}{.7\textwidth}
  \centering
  \includegraphics[width=\linewidth]{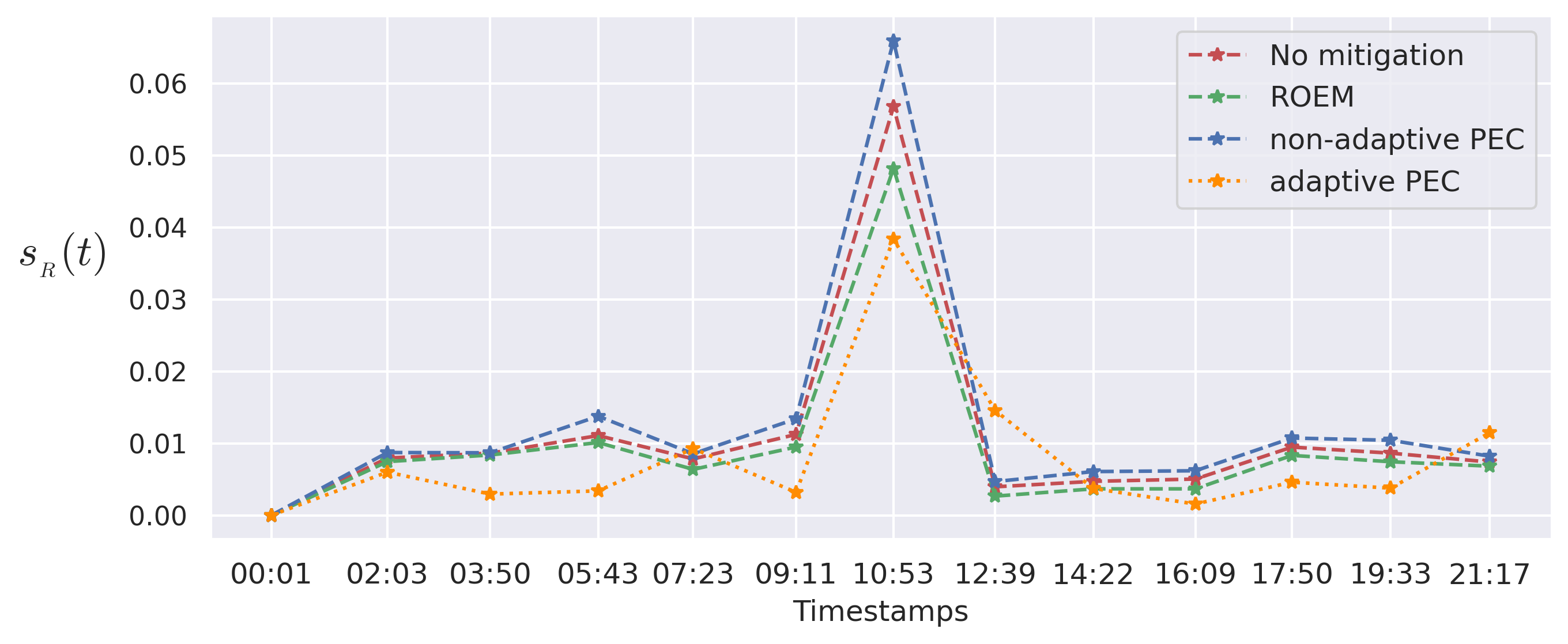}
  \caption*{(b)}
\end{minipage}
\caption{
The figure comprises two plots, each depicting four graphs: (1) "No mitigation" presents raw Bernstein-Vazirani metrics without error mitigation, (2) "ROEM" shows metrics after readout error mitigation, with constant SPAM noise parameters, (3) "Non-adaptive PEC" displays metrics for PEC, and (4) "Adaptive PEC" exhibits metrics for PEC with adaptive noise mitigation. The x-axis denotes intra-calibration time-stamps (UTC) for Jan 15. \added{Plot (a) illustrates the time-varying accuracy metric from Eqn.~\ref{eq:eR}.} It demonstrates the adaptive PEC's 42\% accuracy improvement over non-adaptive PEC. \added{Plot (b) shows the time-varying stability metric from Eqn.~\ref{eq:sR}.} It illustrates the adaptive PEC's 60\% stability enhancement compared to non-adaptive PEC. These plots underscore the significant impact of non-stationary noise on PEC resilience. Due to PEC's lengthy completion time (a couple of hours), adaptive methods are able to handle non-stationary noise conditions better.
}
\label{fig:stability_accuracy}
\end{figure*}
%\FloatBarrier
%%%%%%%%%%%%%%%%%%%%%%%%%%%%%%%%%%%%%%%%%%%%%
\subsection{Non-stationary noise estimates}
Fig.~\ref{fig:tv_bv_depol_spam} illustrates how noise in the quantum computer changed over time. The blue line in plot (a) shows the depolarizing parameter for the target qubit of the CNOT gate, while the black line represents the depolarizing parameter for the control qubit. Plot (b) shows five lines, each representing the SPAM fidelity for the register elements. The x-axis denotes intra-calibration timestamps. 
%The parameters were estimated using the methodology outlined in Section \ref{sec:exp_dataset} and the data was elaborated upon in Section \ref{sec:bayesian-stabilization}.
The graph shows periods where the noise levels in the depolarizing parameter for qubit 3 (the control qubit) are steady, notably between 2:00 am and 7:30 am, contrasting with times of significant fluctuation, as observed between 9:00 am and 2:30 pm. The depolarizing parameter fluctuates between 1\% and 4\% for qubit 4 (the target qubit) and between 1\% and 3\% for qubit 3, both peaking sharply at 10:53 am.

\added{
Since the probabilities of observing the computational basis states on the left-hand side in Eqn.~33 are functions of time, we expect the SPAM fidelities to also vary over time. 
However, from Fig.~6~(b), the value for qubit 3 appears almost constant. 
This qubit exhibits less non-stationarity compared to other qubits, suggesting greater reliability with respect to SPAM noise. 
However, these estimates are influenced by grid discretization used in the mean square error minimization procedure for solving the non-linear equations to find the best-fit. 
Our grid for SPAM fidelity had a spacing of $0.01$ in the range $[0, 1]$. 
Our optimizer consistently produced an estimate for $f_3$ as $0.99$ across all the time-stamps, which is near the upper limit of the permissible range. 
We have not analyzed the impact of discretization granularity on degree of non-stationarity of the solution space.
}

Qubit 4 experiences the most significant impact from SPAM noise, with its values fluctuating between 0.99 and 0.94, a notable range given the sensitivity of PEC  to accurate noise estimations. Meanwhile, Qubit 2 demonstrates a gradual drift in its values, starting from below 0.94 and rising to 0.96. This progressive change suggests a systematic, non-random trend that might be rectifiable with bias shift corrections. However, such patterns are not uniform across the entire register, implying the necessity to consider non-stationary statistics for modeling the system. Conducting experiments in times of significant non-stationary activity, like from 7:30 am to 2:30 pm, lead to more unstable outcomes when using non-adaptive PEC.
\subsection{Non-stationary quasi-probability distribution}
Fig.~\ref{fig:tv_qpr} underscores the importance of considering the \ns~nature of the quasi-probability distribution when implementing PEC, especially given the lengthy data collection process required for a single PEC mitigation (approximately 2 hours in our example). The abrupt change observed at 12:39 p.m. in the quasi-probability distribution directly correlates with the sharp change in the noise parameters characterizing the quantum circuit at the same time, as illustrated in Fig.~\ref{fig:tv_bv_depol_spam}.

To maintain clarity in Fig.~\ref{fig:tv_qpr}, we have not plotted all 512 bins of the histogram in one plot. 
Fig.~\ref{fig:tv_qpr}~(a) displays the time-varying weight $\mathds{Q}_0$ for the Bernstein-Vazirani circuit as-is without any additional Pauli-gate added (we also refer to this as the raw \BV circuit). 
If the circuit were noiseless, then the weight for the raw circuit will be a constant 1. 
We observe a decrease in weight for the raw circuit, dipping below 78\% around 10:53 am from a peak of almost 83\%, coinciding with a peak in circuit noise as seen in Fig.~\ref{fig:tv_bv_depol_spam}. This decrease in weight is expected as the circuit noise peaks.

The values of the quasi-probability bins for the next 10 basis circuits are shown in plot (b), with values approximately 10 times lower than the first circuit. 
%Among these, the basis circuit with index 4 exhibits the most fluctuating value. The values for subsequent basis circuits are even smaller. 
Ignoring seemingly small coefficients in the quasi-probability distribution without considering the precision of final reported results can be risky. Subsequent basis circuits, not shown here, have significantly smaller quasi-probability weights (around $10^{-4}$). Yet their collective impact in a sum of 500 can be substantial, contributing up to 0.05. Given our reported accuracy and stability are around $10^{-2}$, these coefficients, though small, can significantly influence the results. Therefore, in our analysis, we included all basis circuits without approximation, focusing on the effects of non-stationary noise on PEC, rather than on resource optimization.
\subsection{Non-stationary PEC outcomes}
The impact of the non-stationary noise can be seen in Fig.~\ref{fig:stability_accuracy}. The first plot, labeled No mitigation, presents the accuracy and stability metrics defined in Eqns.~\ref{eq:eR} and \ref{eq:sR} respectively, for the raw Bernstein Vazirani circuit without any form of quantum error mitigation. The second plot, labeled ROEM, which stands for readout error mitigation, displays the metrics after performing SPAM noise mitigation. In this case, the SPAM noise parameters are held constant after initial device characterization. It deploys the standard matrix inversion~\cite{bravyi2021mitigating} technique for mitigation. The third plot, labeled non-adaptive PEC, exhibits the accuracy and stability metrics for the Bernstein Vazirani circuit with non-adaptive PEC. This method incorporates SPAM noise mitigation within the PEC framework, as detailed in Sec.~\ref{sec:analytical_BV_model}. The fourth plot, labeled adaptive PEC, presents the metrics for the adaptive PEC method, as discussed in Sec.~\ref{sec:adaptPEC}.

Fig.~\ref{fig:stability_accuracy}~(a) shows the effectiveness of the adaptive PEC in enhancing result accuracy. It reveals approximately a 42\% improvement in accuracy on average compared to the non-adaptive method. The observed accuracy benefit ranges from a minimum of 25\% to a maximum of 78\%.  Fig.~\ref{fig:stability_accuracy}~(b) shows the impact on result stabilization. It shows an approximately 60\% enhancement in stability on average compared to the non-adaptive method. The observed stability benefit ranges from a minimum of 8\% to a maximum of 200\%.

Observing the plots, it is evident that adaptive PEC significantly outperforms standard PEC. Additionally, all four methods (no mitigation, ROEM, PEC, and adaptive PEC) exhibit a time-series trend that deteriorates notably at 10:53 am. The observation correlates with the abrupt change in the underlying quasi-probability distribution at 10:53 am as seen in Fig.~\ref{fig:tv_qpr}. 

Both the accuracy and stability metrics at 12:39 p.m. are slightly worse for adaptive PEC compared to non-adaptive PEC. This discrepancy stands out as the only instance where adaptive PEC performs sub-optimally. It seems, surprisingly, that the stale data serves as a better noise estimate for this specific time-point. However, the noise at 12:39 p.m. is not necessarily closer to the noise at the starting time-stamp of 00:01 a.m., as demonstrated in Fig.~\ref{fig:tv_bv_depol_spam}. While adaptive PEC succeeds in most cases, the learning process is not instantaneous. There exists a slight lag in learning due to the influence of prior information on the final estimate. This phenomenon reflects a fundamental aspect of learning methods: the presence of memory, which can aid learning but also slows down adaptation to the fast, spiky changes. The sharp change in noise at 10:53 a.m. leads to an overestimation of the noise estimate compared to when using the initial value at 00:01 a.m. as a reference point. Residual errors in the noise estimates likely arises from inaccurate models and non-stationary processes changing at a faster rate than sampled here.
\section{Conclusion}\label{sec:conclusion}
Characterizing noise in contemporary quantum devices remains challenging due to their non-stationary statistics~\cite{etxezarreta2021time}, rendering mitigation strategies~\cite{bharti2022noisy} highly susceptible to errors arising from statistical estimation. Incorrect assumptions, such as neglecting non-stationarity, can yield misleading and irreproducible results.

This study examined the impact of non-stationary noise on probabilistic error cancellation (PEC). It introduced a Bayesian approach to improve stability and accuracy of PEC outcomes. We tested our algorithm on the \kolkata device on January 15, 2024. The dataset covered a 24-hour period, consisting of 13 complete PEC datasets for the Bernstein-Vazirani test circuit, with approximately 67 million observations. 
Results indicate a 42\% increase in accuracy and a 60\% enhancement in stability compared to non-adaptive PEC. Consistent improvement trends across time-stamps and qubits provide robust evidence for the effectiveness of adaptive PEC. 

The choice of noise model influences the estimation process. 
If the chosen noise model is incorrect or insufficiently granular, the estimated noise parameters will reflect the specific dataset's inherent patterns rather than accurately representing a true noise model applicable for various inputs. 
Mitigation using such mis-estimated noise models will be unsuccessful because the mitigation algorithm will introduce additional errors to the already noisy data. 
Improvements in mitigation results, including improved performance vs non-adaptive PEC shown here, indicate that the noise is sufficiently well-characterized using our simplified noise model but additional improvements on accuracy and stability can be obtained by using more frequent adaptive estimates. 

\added{
Our goal in this paper was not to provide examples of performing Bayesian inference in different quantum computing contexts. Instead, we aimed to highlight that non-stationary noise is a significant issue in contemporary quantum computers, necessitating adaptive methods. Bayesian inference represents one such method. While our framework is generic, the specific equations for inference will differ based on the application. We illustrated our methodology using \BV as a test case for improving \pec in presence of \nsn.
}

\added{
Before concluding, we will discuss the scalability of the proposed algorithm. There are three potential areas where scalability may be a concern: PEC resource costs, optimization costs, and sampling costs.\\\\
A) PEC resource costs: The first scalability concern pertains to the number of PEC circuits required and is not directly related to Bayesian inference. The problem is well-known: each CNOT gate requires 16 PEC circuits for mitigating CNOT noise, and each qubit needs 2 PEC circuits for SPAM noise mitigation. Consequently, a \BV circuit with \( m \) CNOT gates and \( n \) qubits will need \( 2^{4m+n} \) PEC circuits. To mitigate this, one strategy is to avoid using PEC for SPAM mitigation and instead use standard matrix inversion (which may compromise some accuracy).  Additionally, noise can be modeled in modular circuit layers, with PEC applied to these modules (instead of individual quantum gates).\\\\
B) Optimization: The second concern involves solving the non-linear system of equations (cf. Eqns.~\ref{eq:tv-correlated-means}). In the Bernstein-Vazirani circuit, with \(n\) qubits and \(m\) CNOT gates, there are \(2m + n\) noise parameters: each qubit contributes one SPAM parameter, and each CNOT gate's asymmetric depolarizing noise model contributes \(2m\) parameters. For an \(n\)-qubit problem, there can be \(2^n\) such equations, one for each computational basis state. When \(2m + n < 2^n\), we have an over-determined system that can be solved using mean square error minimization. However, this is computationally expensive due to the high-dimensional search space for correlated parameters. If the system is under-determined, the estimates are unreliable due to multiple possible solutions. To make the algorithm scalable, focus on the most significant sources of noise and avoid overly detailed models that do not contribute significantly to noise explanation. For example, in this paper, we ignored gate noise in single-qubit gates and focused on SPAM noise and noise in entangling gates. In fact, even with quantum error correction, statistical mitigation methods may be relevant due to unforeseen noise sources. A single logical qubit may consist of thousands of physical qubits, so it is crucial to model noise at an aggregate, modular level (e.g., the logical level) rather than at lower levels of the computing stack. Choosing the right level of abstraction and avoiding overly granular noise models reduces the number of noise parameters that need to be estimated.\\\\
C) Sampling: Parameter estimation using Bayesian inference often requires sampling from a multivariate correlated distribution. However, the algorithm presented in this paper for estimating time-varying circuit noise parameters eliminates the need for sampling. As demonstrated in Eqn.~\ref{eq:posterior_dirichlet}, the parameters of the Dirichlet distribution, which models the joint density for the probability simplex, are updated using observed data \(\mathcal{D}\), which help us formulate the system of non-linear equations to estimate the time-varying noise parameters. However, if custom metrics such as 95\% confidence intervals are needed, or if we prefer using the maximum-a-posteriori (MAP) estimate instead of the mean, then Monte Carlo sampling will be required, which has a high computational overhead.
}
\section*{Code and Data}
The code and data are for available publicly available in the github repository: \href{https://github.com/quantumcomputing-lab/pecTVQN/}{https://github.com/quantumcomputing-lab/pecTVQN/}
\section*{Acknowledgments}
This material is based upon work supported by the US Department of Energy, Office of Science, National Quantum Information Science Research Centers, Quantum Science Center.

\bibliographystyle{unsrt}
\bibliography{references.bib}
\end{document}